\documentclass[11pt,a4paper]{article}%
\pdfoutput=1

\usepackage{amsmath}
\usepackage{cite}
\usepackage{graphicx}
\usepackage{bbm}
\usepackage{amsfonts}
\usepackage{amssymb}
\newcommand{\be}{\begin{equation}}
\newcommand{\ee}{\end{equation}}

\newcommand{\SU}[1]{\ensuremath{\mathrm{SU}(#1)}}

\newcommand{\tr}{\operatorname{tr}}

\newcommand{\Y}{{\ensuremath{\bf Y}}}

\newcommand{\trans}{\ensuremath{\mathrm{T}}}

\renewcommand{\Im}{\operatorname{Im}}

\begin{document}

\thispagestyle{empty} 

\vspace*{-10mm}
\begin{flushright}
LPSC14022\\
SISSA  07/2014/FISI
\end{flushright}

\begin{center}

\vspace*{3cm}

\textbf{\LARGE The Flavour of Natural SUSY}

\vspace*{1cm}

{\large Felix Br\"ummer${^1}$, Sabine~Kraml${^2}$, Suchita~Kulkarni${^2}$, Christopher Smith${^2}$}

\vspace*{1cm}

{\it
$^1$~SISSA/ISAS, Via Bonomea 265, I-34136 Trieste, Italy\\[2mm]
$^2$~Laboratoire de Physique Subatomique et de Cosmologie, Universit\'e Grenoble-Alpes, CNRS/IN2P3, 53 avenue des Martyrs, F-38026 Grenoble Cedex, France\\[2mm]
}

\end{center}

\vspace*{1cm}

\begin{abstract}
\noindent
An inverted mass hierarchy in the squark sector, as in so-called ``natural supersymmetry'', requires non-universal boundary conditions at the mediation scale of supersymmetry breaking. We propose a formalism to define such boundary conditions in a basis-independent manner and apply it to generic scenarios where the third-generation squarks are light, while the first two generation squarks are heavy and near-degenerate. We show that not only is our formalism particularly well-suited to study such hierarchical squark mass patterns, but in addition the resulting  soft terms at the TeV scale are manifestly compatible with the principle of minimal flavour violation, and thus automatically obey constraints from flavour physics. 
\end{abstract}

\clearpage
\section{Introduction}

In supersymmetric extensions of the Standard Model (SM), any particles with sizeable couplings to the Higgs sector are expected to have masses not too far above the electroweak scale. This concerns in particular the squarks of the
third generation, which should be lighter than about a TeV in order not to
create a severe naturalness problem. By contrast, the squarks of the first two
generations could well be much heavier.
This possibility is particularly attractive because the bounds from supersymmetry 
(SUSY) searches at the LHC are strongest by far for the first two generation 
squarks, and because flavour constraints are also easier to satisfy when they are 
very heavy. 
The scenario of an inverted mass hierarchy in the squark sector, typically combined with a small higgsino mass parameter and a not too heavy gluino (see {\it e.g.}~\cite{Cohen:1996vb,Kitano:2006gv,Barbieri:2009ev,Papucci:2011wy} and references therein), is commonly dubbed ``natural'' or ``effective''  
SUSY, and is increasingly becoming the new paradigm of SUSY phenomenology.  
 
In the Minimal Supersymmetric Standard Model (MSSM) with boundary conditions 
at the Grand Unification (GUT) scale, 
light stops and sbottoms with otherwise very heavy squarks are especially 
interesting because they can lead to radiatively induced large stop mixing
\cite{Baer:2012uy,Brummer:2012ns,Badziak:2012rf}. The latter is needed in the MSSM to obtain a 126~GeV Higgs mass while keeping the stops reasonably light.
More precisely, if the first two generation squarks have masses of the order
of $10$ TeV, and if supersymmetry breaking is mediated at a very high scale
such as $M_{\mathrm{GUT}}\approx10^{16}$ GeV, then the stop masses at the low
scale receive significant negative contributions from two-loop running (or
possibly even from one-loop running if there is a non-vanishing hypercharge $D$-term). This allows to realize a sizeable ratio $\left|  A_{t}/\overline{m}_{\tilde t}\right|$,
where $A_{t}$ is the stop trilinear parameter and $\overline{m}_{\tilde t}$ is the
average stop mass, leading to large one-loop corrections to the lightest Higgs
mass. However, in precisely this situation where radiative corrections to the
spectrum from the first two generations are important, they may also induce a
significant misalignment between the squark and quark mass matrices. The
resulting flavour-changing neutral currents (FCNCs) are tightly constrained by 
experiment. The effects of such a split squark spectrum on flavour observables have already been investigated in \cite{Giudice:2008uk, Kersten:2012ed, Mescia:2012fg, Barbieri:2010ar} (see also \cite{Gherghetta:2011wc,Auzzi:2011eu, Larsen:2012rq, Craig:2012di, Eliaz:2013aaa, Dudas:2013gga, Arvanitaki:2013yja,Brummer:2013upa,Dudas:2013pja} for some recent discussions on FCNCs in selected models with light third generation squarks). Here, we propose to shed light on this issue using a different strategy.

Firstly, having assumed a very high mediation scale, hierarchical squark soft terms at the low scale have to be obtained from some non-universal boundary conditions through the renormalization group evolution. But even just prescribing such boundary conditions in a model-independent way is nontrivial, since they depend on the chosen flavour basis. Our first result is to propose a formalism to define general soft term boundary conditions in a basis-independent manner. Secondly, we apply this formalism to the cases where either a subset or all of the third-generation squarks are light, while the first two generation squarks are heavy and near-degenerate. It turns out that not only is our formalism particularly well-suited to study such squark mass patterns, but in addition the resulting TeV-scale soft terms are in 
many cases manifestly compatible with the minimal flavour violation principle (MFV)\footnote{In this respect, it should be stressed that our strategy to prove the compatibility of hierarchical 
squark mass patterns with FCNC constraints is different from that of Ref.~\cite{Barbieri:2010ar}, which relied on a modified MFV principle, based on a smaller flavour symmetry group.}, as proposed in Ref.~\cite{D'Ambrosio:2002ex}. In addition, whenever a departure from MFV is observed, it can be quantified precisely. Clearly, realizing split squark scenarios in this way is of great advantage because it helps ensure that there will be no conflict with bounds on $D-\bar D$ and $K-\bar K$ mixing observables, which one might otherwise expect for generic hierarchical soft terms. 

In Section~2, we briefly recall the essentials of the SUSY flavour problem, the concept of MFV, and present our procedure to define fully generic and non-universal boundary conditions for soft-breaking terms. In Section~3 we use this scheme to parametrize the boundary conditions leading to third-generation squarks much lighter than the first two generations, and characterize their flavour properties. Section~4 contains our conclusions. In the appendix, we address some technical subtleties regarding the definition and running of the CKM matrix and show that our scheme allows to easily deal with, and correct for, CKM-induced uncertainties in the renormalization group (RG) running.

\section{The SUSY flavour sector}

We follow the conventions of the SLHA2 \cite{Allanach:2008qq}, which we now briefly recall.
The matter fields of the supersymmetric Standard Model transform under
a global non-abelian flavour symmetry $G_{\rm F}=\mathrm{SU}(3)_{Q}\times
\mathrm{SU}(3)_{U}\times\mathrm{SU}(3)_{D}\times\mathrm{SU}(3)_{L}%
\times\mathrm{SU}(3)_{E}$. This symmetry is explicitly broken by the Yukawa
superpotential
\begin{equation}
W_{\mathrm{Yukawa}}=-(\mathbf{Y}_{u})_{ij}\,H_{u}Q_{i}U_{j}+(\mathbf{Y}%
_{d})_{ij}\,H_{d}Q_{i}D_{j}+(\mathbf{Y}_{e})_{ij}\,H_{d}L_{i}E_{j}\;,
\label{WYukawa}
\end{equation}
as well as by the soft mass matrices for the squarks and sleptons, and by the
soft trilinear terms.

In the lepton-slepton sector, $\mathbf{Y}_{e}$ can always be diagonalized via a suitable
$\mathrm{SU}(3)_{L}\times\mathrm{SU}(3)_{E}$ transformation. We will 
focus on the quark-squark sector, where at most one of the matrices $\mathbf{Y}_{u}$
and $\mathbf{Y}_{d}$ can be chosen diagonal in a gauge eigenstate basis.
After electroweak symmetry breaking, the Yukawa matrices are diagonalized by%
\begin{equation}
\widehat{\mathbf{Y}}_{d,u}=\mathbf{V}_{d,u}^{R\dag}\mathbf{Y}_{d,u}%
^{\mathrm{T}}\mathbf{V}_{d,u}^{L}\,,\qquad\qquad\widehat{\mathbf{Y}}%
_{d}=\text{diag}(y_{d},\,y_{s},\,y_{b})\,,\qquad\widehat{\mathbf{Y}}%
_{u}=\text{diag}(y_{u},\,y_{c},\,y_{t})\,\;.\label{yukdiag}%
\end{equation}
The misalignment of left-handed quarks is encoded in the CKM matrix,
$\mathbf{V}_{\mathrm{CKM}}=\mathbf{V}_{u}^{L\dag}\mathbf{V}_{d}^{L}$. Rotating
quarks and squarks by the same unitary transformations defines the super-CKM
basis, in which the squark mass matrices take the form
\begin{equation}%
\begin{split}
\mathbf{M}_{\tilde{u}}^{2}= &  \left(
\begin{array}
[c]{cc}%
{\mathbf{V}_{\mathrm{CKM}}\,\widehat{\mathbf{m}}}_{Q}^{2}\mathbf{V}%
_{\mathrm{CKM}}^{\dag}+\frac{v_{u}^{2}}{2}\widehat{\mathbf{Y}}_{u}^{2} &
\frac{v_{u}}{\sqrt{2}}\left(  \widehat{\mathbf{T}}_{u}^{\dag}-\widehat
{\mathbf{Y}}_{u}\,\mu\cot\beta\right)  \\
\frac{v_{u}}{\sqrt{2}}\left(  \widehat{\mathbf{T}}_{u}-\widehat{\mathbf{Y}%
}_{u}\,\mu^{\ast}\cot\beta\right)   & \widehat{\mathbf{m}}_{U}^{2}%
+\frac{v_{u}^{2}}{2}\widehat{\mathbf{Y}}_{u}^{2}%
\end{array}
\right)  +D\text{-terms}\,,\\
\mathbf{M}_{\tilde{d}}^{2}= &  \left(
\begin{array}
[c]{cc}%
{\widehat{\mathbf{m}}}_{Q}^{2}+\frac{v_{d}^{2}}{2}\widehat{\mathbf{Y}}_{d}^{2}
& \frac{v_{d}}{\sqrt{2}}\left(  \widehat{\mathbf{T}}_{d}^{\dag}-\widehat
{\mathbf{Y}}_{d}\,\mu\tan\beta\right)  \\
\frac{v_{d}}{\sqrt{2}}\left(  \widehat{\mathbf{T}}_{d}-\widehat{\mathbf{Y}%
}_{d}\,\mu^{\ast}\tan\beta\right)   & \widehat{\mathbf{m}}_{D}^{2}%
+\frac{v_{d}^{2}}{2}\widehat{\mathbf{Y}}_{d}^{2}%
\end{array}
\right)  +D\text{-terms}\,.
\end{split}
\label{eq:squarkmasses}%
\end{equation}
In terms of the interaction-basis soft masses $\mathbf{m}_{Q,U,D}^{2}$ and
trilinear terms $\mathbf{T}_{u,d}$,
\begin{align}
\widehat{\mathbf{m}}_{Q}^{2} &  =\,\mathbf{V}_{d}^{L\dag}\mathbf{m}_{Q}%
^{2}\mathbf{V}_{d}^{L}\,,\quad\widehat{\mathbf{m}}_{U}^{2}=\mathbf{V}%
_{u}^{R\dag}(\mathbf{m}_{U}^{2})^{\mathrm{T}}\mathbf{V}_{u}^{R}\,,\quad
\widehat{\mathbf{m}}_{D}^{2}=\mathbf{V}_{d}^{R\dag}(\mathbf{m}_{D}%
^{2})^{\mathrm{T}}\mathbf{V}_{d}^{R}\,,\;\\
\widehat{\mathbf{T}}_{u} &  =\,\mathbf{V}_{u}^{R\dag}\mathbf{T}_{u}%
^{\mathrm{T}}\mathbf{V}_{u}^{L}\,,\quad\widehat{\mathbf{T}}_{d}=\mathbf{V}%
_{d}^{R\dag}\mathbf{T}_{d}^{\mathrm{T}}\mathbf{V}_{d}^{L}\,\;. \nonumber
\end{align}

Our aim is now to establish a formalism for encoding the squark sector soft term
data without fixing a flavour basis. Such a basis-independent formalism has both
conceptual and practical advantages which will be discussed in detail below.

In order to find a basis-independent parameterization of the soft terms, we expand
them in powers of the Yukawa matrices, covariantly with respect to the spurious
$G_{\rm F}$ flavour symmetry. To this end we define the matrices
\begin{equation}
\mathbf{A}=\mathbf{Y}_{d}\mathbf{Y}_{d}^{\dag}\,\;,\qquad\mathbf{B}%
=\mathbf{Y}_{u}\mathbf{Y}_{u}^{\dag}\;.
\end{equation}
They transform as bifundamentals under an $\mathrm{SU}(3)_{Q}$ rotation 
which sends
$Q\rightarrow\mathbf{V}_{Q}Q$:%
\begin{equation}
\mathbf{A}\rightarrow\mathbf{V}_{Q}^{\ast}\mathbf{A}\,\mathbf{V}%
_{Q}^{\mathrm{T}}\,\;,\qquad\mathbf{B}\rightarrow\mathbf{\mathbf{V}}_{Q}%
^{\ast}\mathbf{B}\,\mathbf{V}_{Q}^{\mathrm{T}}\,\;.
\end{equation}
Given that $\mathbf{m}_{Q}^{2}$ also transforms as a bifundamental, $(\mathbf{m}%
_{Q}^{2})^{\mathrm{T}}\rightarrow\mathbf{V}_{Q}^{\ast}(\mathbf{m}_{Q}%
^{2})^{\mathrm{T}}\,\mathbf{V}_{Q}^{\mathrm{T}}$, we can expand
\begin{equation}\begin{split}\label{Expand1}
\left({\bf m}_Q^2\right)^\trans=m_0^2\Bigl(a_1^q\,{\bf 1}+a_2^q\,{\bf A}&+a_3^q\,{\bf B}+a_4^q\,{\bf A}^2+a_5^q\,{\bf B}^2 + a_6^q\,\{{\bf A},{\bf B}\}\\
 &+i\,b_1^q\, [{\bf A},{\bf B}] +i\,b_2^q\, [{\bf A},{\bf B}^2]+i\,b_3^q\, [{\bf B},{\bf A}^2]\Bigr)\,,
\end{split} 
\end{equation}
where the expansion coefficients $a_i^q$ and $b_i^q$ are invariant under $G_{\rm F}$.
Likewise, given their respective transformation properties under $G_{\rm F}$, the right-handed
squark mass matrices and the trilinear terms are covariantly expanded as
\be\begin{split}\label{Expand2}
{\bf m}_{U}^2=m_0^2\Bigl(a_1^{u}\,{\bf 1}+{\bf Y}_{u}^\dag\bigl(a_2^{u}\,{\bf 1}+a_3^{u}\,{\bf A}&+a_4^{u}\,{\bf B}+a_5^{u}{\bf A}^2+ a_6^{u}\,\{{\bf A},{\bf B}\} \\
 &+i\,b_1^{u}\, [{\bf A},{\bf B}] +i\,b_2^{u}\, [{\bf A},{\bf B}^2]+i\,b_3^{u}\, [{\bf B},{\bf A}^2]\bigr){\bf Y}_{u}\Bigr)\,,\\
{\bf m}_{D}^2=m_0^2\Bigl(a_1^{d}\,{\bf 1}+{\bf Y}_{d}^\dag\bigl(a_2^{d}\,{\bf 1}+a_3^{d}\,{\bf A}&+a_4^{d}\,{\bf B}+a_5^{d}\,{\bf B}^2+a_6^{d}\,\{{\bf A},{\bf B}\} \\
 &+i\,b_1^{d}\, [{\bf A},{\bf B}] +i\,b_2^{d}\, [{\bf A},{\bf B}^2]+i\,b_3^{d}\, [{\bf B},{\bf A}^2]\bigr){\bf Y}_{d}\Bigr)\,,
\end{split} 
\ee
\be\begin{split}\label{Expand3}
{\bf T}_{u,d}=A_0 \Bigl(c_1^{u,d}\,{\bf 1}+c_2^{u,d}\,{\bf A}&+c_3^{u,d}\,{\bf B}+c_4^{u,d}\,{\bf A}^2+c_5^{u,d}\,{\bf B}^2+c_6^{u,d}\,\{{\bf A},{\bf B}\} \\
 &+i\,c_7^{u,d}\, [{\bf A},{\bf B}] +i\,c_8^{u,d}\, [{\bf A},{\bf B}^2]+i\,c_9^{u,d}\, [{\bf B},{\bf A}^2]\Bigr){\bf Y}_{u,d}\,.
\end{split} 
\ee
The coefficients $a_{i}^{q,u,d}$, $b_{i}^{q,u,d}$ are real because the mass matrices are
hermitian, but the $c_{i}^{u,d}$ are generally complex. The parameters $m_0$ and $A_0$
are placeholder constants of mass dimension one which could as well be absorbed into the $a$, $b$,
and $c$ coefficients at one's convenience. Eqs.~\eqref{Expand1}--\eqref{Expand3}
define our basis-independent general parameterization of the squark sector soft terms.

Since the matrices appearing on the RHS  of Eq.~\eqref{Expand1} are linearly independent
(for generic ${\bf A}$ and ${\bf B}$) \cite{Nikolidakis08}, there is no loss of generality
in this expansion. The same is true for each of Eqs.~\eqref{Expand2} and \eqref{Expand3}.
Indeed it is a simple exercise in counting to show
that the real $a_i^{q,u,d}$ and $b_i^{q,u,d}$ together with the complex $c_i^{u,d}$ coefficients contain exactly
the degrees of freedom needed for describing three hermitian $3\times 3$ mass matrices and two
general complex $3\times 3$ trilinear matrices. The bases of flavour-covariant $3\times 3$ matrices we
are projecting on are not unique, but they are in a sense the simplest choices, being symmetric in
$\mathbf{Y}_u$ and $\mathbf{Y}_d$ and using the lowest powers of Yukawa matrices
possible.

These matrix bases turn out to be numerically somewhat peculiar when realistic values for 
${\bf Y}_u$ and ${\bf Y}_d$ are inserted. Because of the large hierarchy 
in the Yukawa couplings, one has $\mathbf{B}^2\approx\tr(\mathbf{B})\mathbf{B}$ and
 $\mathbf{A}^2\approx\tr(\mathbf{A})\mathbf{A}$; that is, some of the basis
matrices are nearly parallel in flavour space. In addition, the only non-diagonal 
structure provided by $\mathbf{A}$ and $\mathbf{B}$ is the very hierarchical 
CKM matrix. Therefore, numerically expanding a generic $3\times3$ matrix 
requires coefficients spanning several orders of magnitude, typically 
up to the order of $m_{t}^{2}/m_{u}^{2}\sim10^{10}$.

The above expansion enables us to adopt a very simple and clear definition
of Minimal Flavour Violation (MFV). The basic assumption of MFV is often stated as 
$G_{\rm F}$ being broken only through powers of Yukawa matrices \cite{D'Ambrosio:2002ex} 
(see also e.g.~\cite{Hall:1990ac,Ali:1999we,Isidori:2006qy}). The usual rationale
is that $G_{\rm F}$ could be an exact but spontaneously broken symmetry
of some more fundamental theory whose dynamics is responsible for the generation of both
the Yukawa couplings and the soft terms. In our framework, we define MFV as follows:
all $a_i^x$, $b_i^x$ and $c_i^x$ coefficients in Eqs.~\eqref{Expand1}--\eqref{Expand3}
should be at most ${\cal O}(1)$ when $m_0$ and $A_0$ represent the typical soft mass scale.
(In fact the statement ``the only sources of $G_{\rm F}$ breaking are powers of Yukawa 
matrices'' is somewhat meaningless when taken on its own, since the above expansion shows 
that one can parameterize any general soft mass and trilinear matrices in this way. However, if
the expansion coefficients are allowed to be arbitrarily large, they could not possibly
originate from  $G_{\rm F}$ spurions in a weakly coupled theory.) For more details, see also
\cite{Mercolli:2009ns, Smith:2009hj}.

At this point we should emphasize that our approach does not rely on the $G_{\rm F}$
symmetry being in any way fundamental. When we allude to MFV in the following, it is mostly
because the MFV condition (in the strict above sense) has certain other appealing properties:
Firstly, it is stable and generally even IR-attractive \cite{Paradisi:2008qh,Colangelo:2008qp} under
the renormalization group; secondly, it allows a model to automatically satisfy many
stringent bounds from flavour physics.

Unification may impose additional relations between the soft terms and hence between
the expansion coefficients. GUT relations are typically spoiled at the subleading
level by higher-dimensional operators involving GUT-breaking VEVs (for instance, the
$\SU{5}$ relation $\Y_d=\Y_e$ should be violated to obtain a valid fermion spectrum).
Neglecting such GUT-breaking effects, one may look for simple conditions on the coefficients
to ensure that the soft terms are compatible with grand unification, depending on
the actual GUT model. For example, standard
$\SU{5}$ unification requires ${\bf m}_Q^2={\bf m}_U^2$. Choosing a basis in which $\Y_u$ 
is diagonal, it is clear that for this to hold it is sufficient to choose
 $a_1^q=a_1^u$, $a_3^q=a_2^u$, $a_5^q=a_4^u$ with all other $a_i^{q,u}=0$. More general
patterns are of course possible since our parametrization is fully general, but they will
in general not be MFV-like.

In this work we are interested in models where the soft terms are neither universal nor necessarily
MFV-like at some very high mediation scale $M_{\rm GUT}\approx 10^{16}$ GeV. We will define the
soft term boundary conditions through the expansion Eqs.~\eqref{Expand1}--\eqref{Expand3}. Such a
procedure has many desirable features (below, we use the short-hand
$x^{q,u,d}=a^{q,u,d},b^{q,u,d},c^{u,d}$; $x=x^{q},x^{u},x^{d}$):

\begin{enumerate}
\item The soft masses and trilinear terms at any scale $Q$ admit 
expansions of
the form~(\ref{Expand1})--(\ref{Expand3}), where both soft terms and Yukawa
couplings are understood as those at the scale $Q$. Thus, the running of the
soft masses and trilinear terms can be represented by that of the flavour
coefficients. Their renormalization group equations (RGEs) were studied in
Refs.~\cite{Paradisi:2008qh,Colangelo:2008qp}. Typically, not only are the
evolutions of the coefficients $a_{i\neq1}[Q], b_i[Q], c_{i\neq1}[Q]$ from $Q=M_{\mathrm{GUT}}$ down to the TeV scale smooth and bounded, but they even exhibit infrared ``quasi''-fixed points, whose values
mostly depend on the non-flavoured MSSM parameters.

\item The $\beta$-functions of the soft masses and trilinear terms are
naturally compatible with the expansions~(\ref{Expand1})--(\ref{Expand3}), and
the running of the various coefficients sum up different physical effects. For
example, the leading coefficients $a_{1}^{q,u,d}[Q],c_{1}^{u,d}[Q]$ entirely
encode the dominant flavour blind evolution, while subleading terms evolve separately.

\item The phenomenological impact of the flavour mixing induced by the
off-diagonal soft term entries can immediately be assessed. Indeed, the
MFV limit is recovered when all the coefficients are $\mathcal{O}(1)$. This
means that one can directly spot potentially dangerous sources of new FCNCs
simply by looking at the relative sizes of the coefficients. For example, if
$a_{1}^{q}[1\,$TeV$]=1$ but $a_{3}^{q}[1$\thinspace TeV$]=1000$, then one
should expect difficulties with FCNC constraints from $K$ and $B$ physics. Indeed, assuming SUSY masses of the order of 1 TeV, such values grossly violate current bounds on mass insertions, see {\it e.g.}~Ref.~\cite{oai:arXiv.org:1002.0900}, with for example $[{\bf m}_Q^2]_{12}/[{\bf m}_Q^2]_{11}\approx 1000\times V_{td}^{*}V_{ts}\sim \mathcal{O}(0.1)$.

\item Starting with universal mSUGRA-like soft-breaking terms, $x_{i}%
[M_{\mathrm{GUT}}]=\delta_{i1}$, the coefficients at the low scale are all
MFV-like, $x_{i}[Q]\sim\mathcal{O}(1)$ or smaller. More generally, the
logarithmic running with small coupling constants cannot upset initial MFV-like
boundary conditions at the GUT scale. The converse is not true though, because
of the presence of the aforementioned quasi-fixed points~\cite{Colangelo:2008qp}.

\item Intrinsically new CP-violating phases, entering exclusively through
$b_{i}^{q,u,d}[M_{\mathrm{GUT}}]\neq0$ and $\operatorname{Im}c_{i}%
^{u,d}[M_{\mathrm{GUT}}]\neq0$, can be simply factored out from CP-violating
effects induced by the CKM phase, introduced through $\mathbf{Y}_{u}$ and
$\mathbf{Y}_{d}$. Note that if $b_{i}^{q,u,d}[M_{\mathrm{GUT}}]=0$ and
$\operatorname{Im}c_{i}^{u,d}[M_{\mathrm{GUT}}]=0$, their values at the
electroweak scale are entirely induced by the CP-violating phase of
$\mathbf{V}_{\mathrm{CKM}}$, and end up tiny. In this respect, the
CP-violating phases of $c_{1}^{u}[Q]$ and $c_{1}^{d}[Q]$ are a bit special.
Being flavour blind, they should be considered along with those of the other
flavour blind complex parameters of the MSSM such as $\mu$ and the gaugino
masses~\cite{Mercolli:2009ns}.

\item For a given boundary condition, one can easily and completely probe its
CKM neighbourhood by allowing $\mathcal{O}(1)$ variations of the coefficients.
Indeed, these variations simulate the presence of arbitrary CKM-like mixing
matrices in both the left ($\mathbf{V}_{u,d}^{L}$) and right-handed sector
($\mathbf{V}_{u,d}^{R}$). In practice, this is far less demanding than it
seems. For $\mathcal{O}(1)$ perturbations, not all the 63 coefficients are
equally relevant, so varying only the first few in each expansion is sufficient.

\item As analysed in the appendix, provided none of the leading coefficients are particularly large, the soft-term expansions are largely independent of the precise parametrization of the CKM matrix. In particular, the coefficients are similar using the full CKM matrix or its CP-conserving limit, no matter how this limit is taken. By contrast, off-diagonal entries of the soft terms can deviate by tens of percent depending on the chosen CKM matrix. This observation is useful in practice since it permits to compute the coefficients under some simplifying assumptions (CP-limit, no threshold corrections, and/or no experimental errors for the CKM parameters), and then to reconstruct with an excellent accuracy the physical soft terms and thereby reliably compute all the flavour observables.

\item Last but not least, it is easy and straightforward to parametrize
boundary conditions where the third-generation squarks are split from the
first two generations, 
since $\mathbf{Y}_{u}\mathbf{Y}_{u}^{\dagger}$ and $\mathbf{Y}_{d}\mathbf{Y}_{d}^{\dagger
}$ do have precisely such a hierarchy. This possibility will be
explored in detail in the next section. 
\end{enumerate}

To be complete, we should point out that there is one practical issue
that needs to be kept in mind. Since the basis matrices span several orders of
magnitude and are approximately linearly dependent, it is necessary to maintain
a high level of accuracy in the numerical evaluations, otherwise instabilities 
can easily arise. This is especially true when computing the coefficients of
highly suppressed terms such as $a_{6}^{u,d}$ or $b_{3}^{u,d}$. For the same 
reason, a perfectly unitary representation of the CKM matrix must be used, 
otherwise spuriously large coefficients can arise.

\section{Split squarks and MFV}

The peculiar structure of the MSSM Yukawa couplings should have its origin
in some unknown flavoured dynamics at some high scale $M_{\mathrm{F}}$. 
If supersymmetry breaking is mediated at a scale greater than $M_{\mathrm{F}}$, then one can
reasonably expect that this flavour dynamics will also generate some non-trivial 
flavour structures for the soft mass terms and the trilinear couplings.
In that sense, expressing the soft terms directly in terms of the Yukawa
couplings through the expansions~(\ref{Expand1})--(\ref{Expand3}) can be regarded
as an attempt at capturing the relationships between them.  
If this picture is correct, the
expansion coefficients at the scale $M_{\mathrm{F}}$ would not be random, but
would derive from the flavour dynamics at that scale. It is thus quite possible
that the various coefficients would actually follow a very definite pattern.

With the above idea in mind, our goal is to design flavour structures leading
to spectra with light third-generation squarks at the low scale. There are many 
ways to achieve this. A first possibility is to impose 
\be
a_3^q\simeq - a_1^q/\langle\mathbf{B}\rangle\,,\qquad a_{i\neq 1,3}^q=b_i^q=0\,,
\ee
where $\langle\mathbf{\cdot}\rangle$ denotes the trace in flavour space. More explicitly, let us set
\begin{align}
\mathbf{m}_{Q}^{2}[M_{\mathrm{GUT}}]  & =m_{0}^{2}\bigl( \mathbf{1}-\alpha
_{q}\mathbf{Y}_{u}\mathbf{Y}_{u}^{\dagger}\langle\mathbf{Y}_{u}\mathbf{Y}%
_{u}^{\dagger}\rangle^{-1}\bigr) ^{\mathrm{T}}\;,\; \nonumber \\
\mathbf{m}_{U,D}^{2}[M_{\mathrm{GUT}}]  & =m_{0}^{2}\mathbf{1}\;,\; \label{eq:scenario1}
\\
\mathbf{T}_{u,d}[M_{\mathrm{GUT}}]  & =A_{0}\mathbf{Y}_{u,d}\;. \nonumber
\end{align}
When the
free parameter $\alpha_{q}$ is close to one, in the basis where $\mathbf{Y}%
_{u}$ is diagonal, $\mathbf{m}_{Q}^{2}$ has its first two entries nearly
degenerate and much larger than the third, which is precisely what we 
aim for. Note, however, that in this particular case the value of $(\mathbf{m}_{Q}^2)_{33}$ 
receives large negative loop corrections from $(\mathbf{m}_{U}^2)_{33}$. In order to 
generate a realistic spectrum, the GUT-scale
$(\mathbf{m}_{Q}^2)_{33}$ cannot not be chosen too small, and/or sizeable positive
corrections from the gaugino masses are needed to overcome this effect. 
At the low scale $\tilde t_L$ and $\tilde b_L$ then end up much lighter
than all the other squarks.

It should be remarked that compared to naively setting%
\begin{equation}
\mathbf{m}_{Q}^{2}[M_{\mathrm{GUT}}]=\left(
\begin{array}
[c]{ccc}%
m_{1}^{2} & 0 & 0\\
0 & m_{1}^{2} & 0\\
0 & 0 & m_{2}^{2}%
\end{array}
\right)  \;,\label{Eq.Usual}%
\end{equation}
our procedure requires the same number of free parameters. But, at the same
time, setting the initial conditions in our way is entirely independent of the
flavour basis, while Eq.~(\ref{Eq.Usual}) in principle requires one to specify
also the four mixing matrices $\mathbf{V}_{u,d}^{L,R}$. In addition, the
parameter $\alpha_{q}$ could bear some physical meaning. First, because
$\langle\mathbf{Y}_{u}\mathbf{Y}_{u}^{\dagger}\rangle^{-1}$ is factored out,
its RG evolution is very flat over the whole range down to the electroweak
scale. Typically $\alpha_q$ changes by $\lesssim 20\%$
during the evolution.  (We explicitly show the evolution of $\alpha_q$
for a different scenario in the following discussion, see Fig.~\ref{fig:alpha}.)
Second, it is tempting to imagine that some 
unknown flavour dynamics sets $\alpha_{q}$ to exactly one at the 
scale $M_{\mathrm{F}}$. 
However, since $M_{\mathrm{F}}\neq M_{\mathrm{GUT}}$, one would then have $\alpha
_{q}[M_{\mathrm{GUT}}]$ close but not exactly equal to one. Thus, the only
phenomenological constraint on this parameter is for it to evolve down to
a value smaller than one at the low scale, so as to avoid inducing negative
eigenvalues for the stop or sbottom squarks and the ensuing colour symmetry breaking.
We are however not aware of any specific flavour model which predicts $\alpha_q=1$,
so for the moment we will treat  $\alpha_q\approx 1$ merely as a parameter choice, 
and study its implications independently of a possible dynamical generation.

A very interesting feature of the boundary condition Eqs.~\eqref{eq:scenario1} 
is that even if left squark masses are highly hierarchical, it nevertheless respects 
the MFV principle since $\langle\mathbf{Y}_{u}\mathbf{Y}_{u}^{\dagger}\rangle\approx
y_{t}^{2}$ is of $\mathcal{O}(1)$ at all scales. So, once evolved to the low
scale, we can immediately predict that these initial conditions should be
compatible with flavour constraints.

\begin{figure}[t!]\centering
\includegraphics[width=0.44\textwidth,clip]{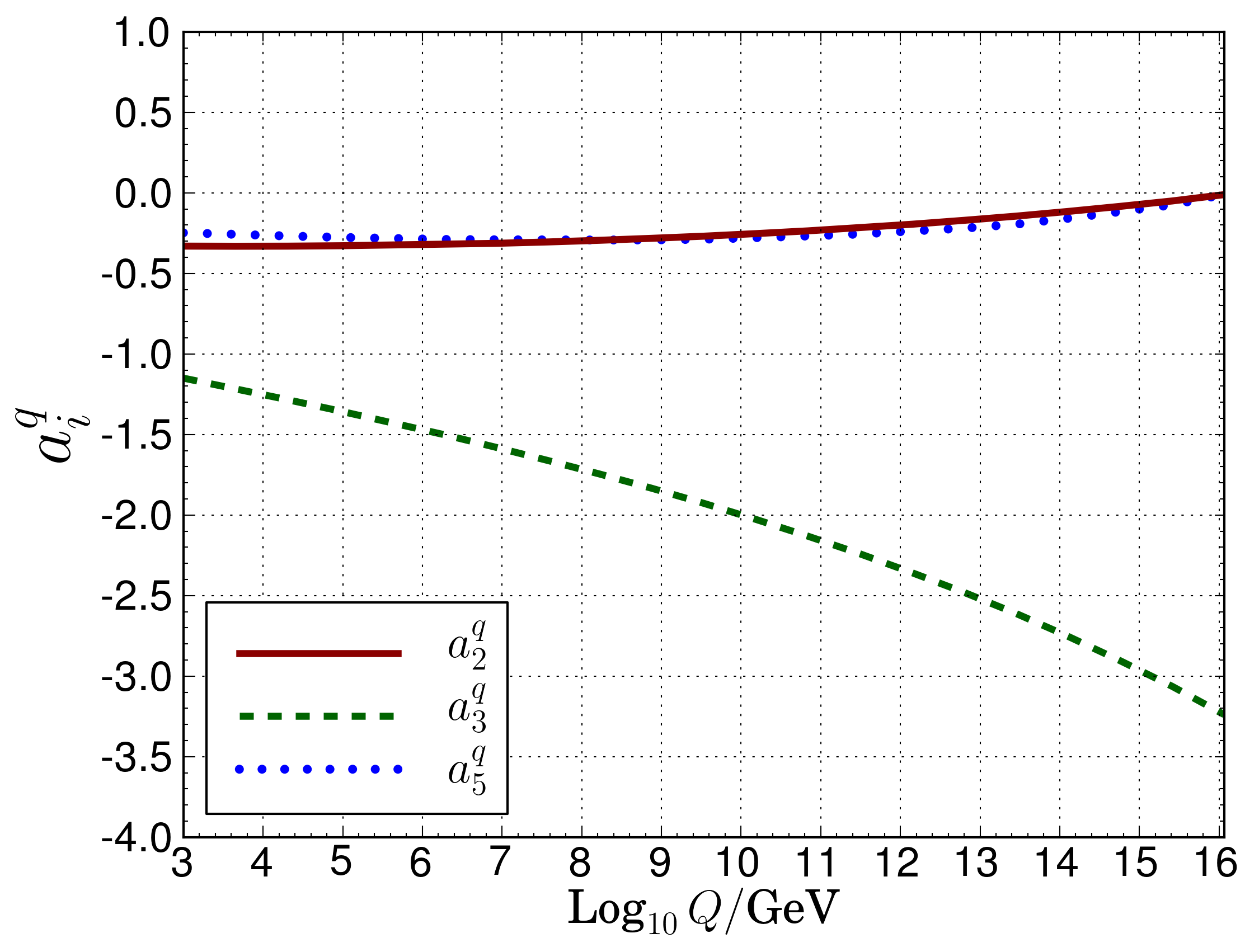}\qquad
\includegraphics[width=0.44\textwidth,clip]{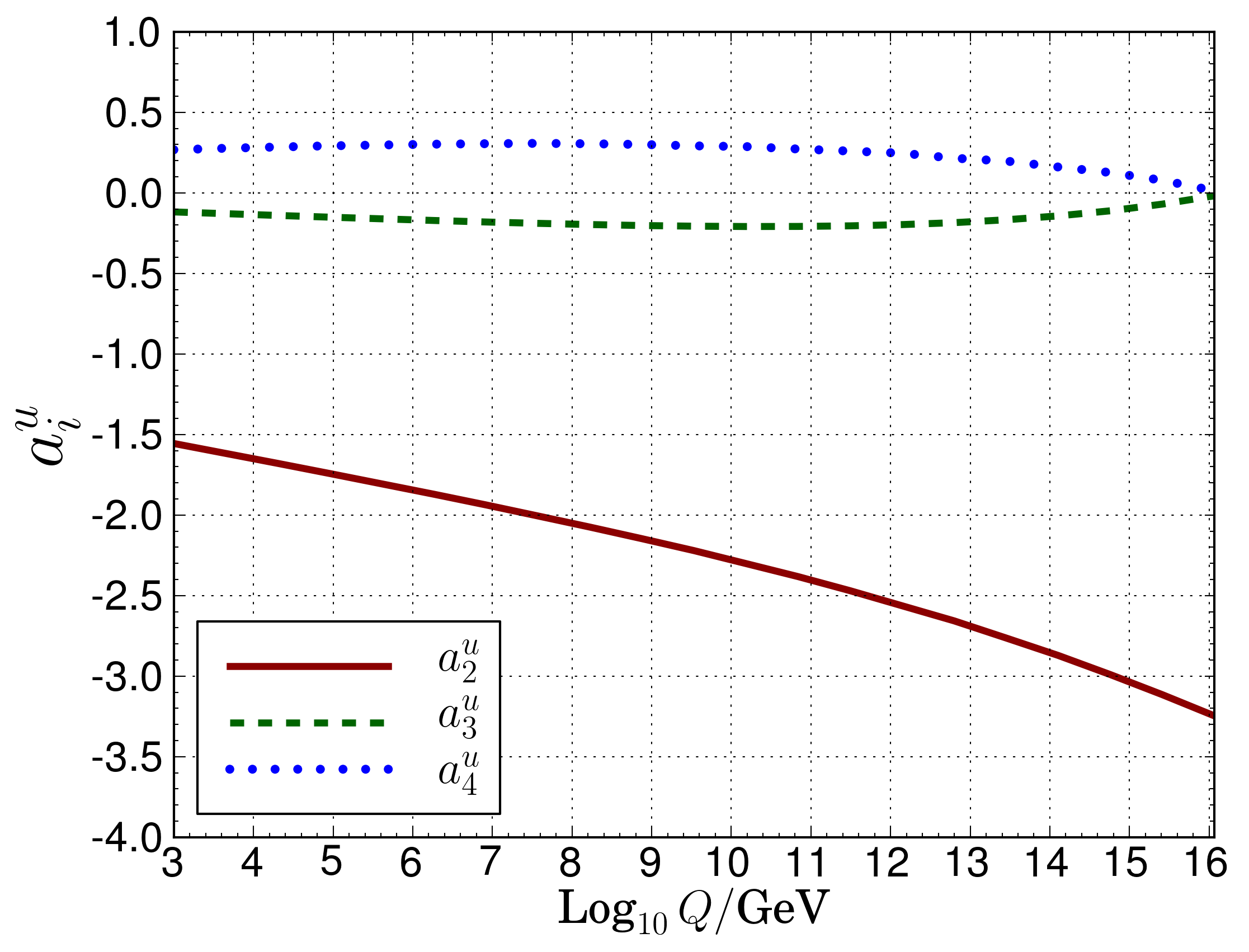}\\
\includegraphics[width=0.44\textwidth,clip]{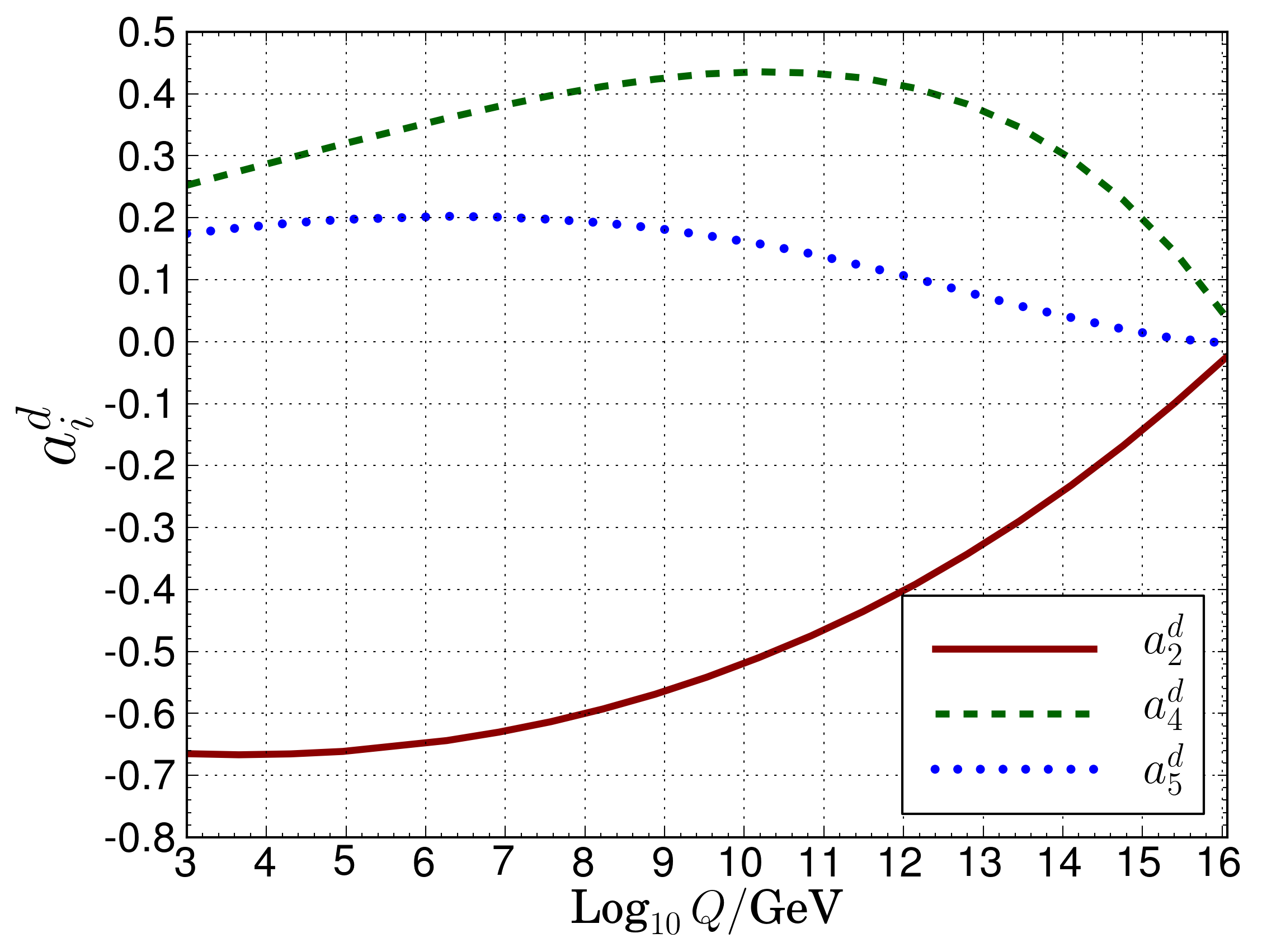}\qquad
\includegraphics[width=0.44\textwidth,clip]{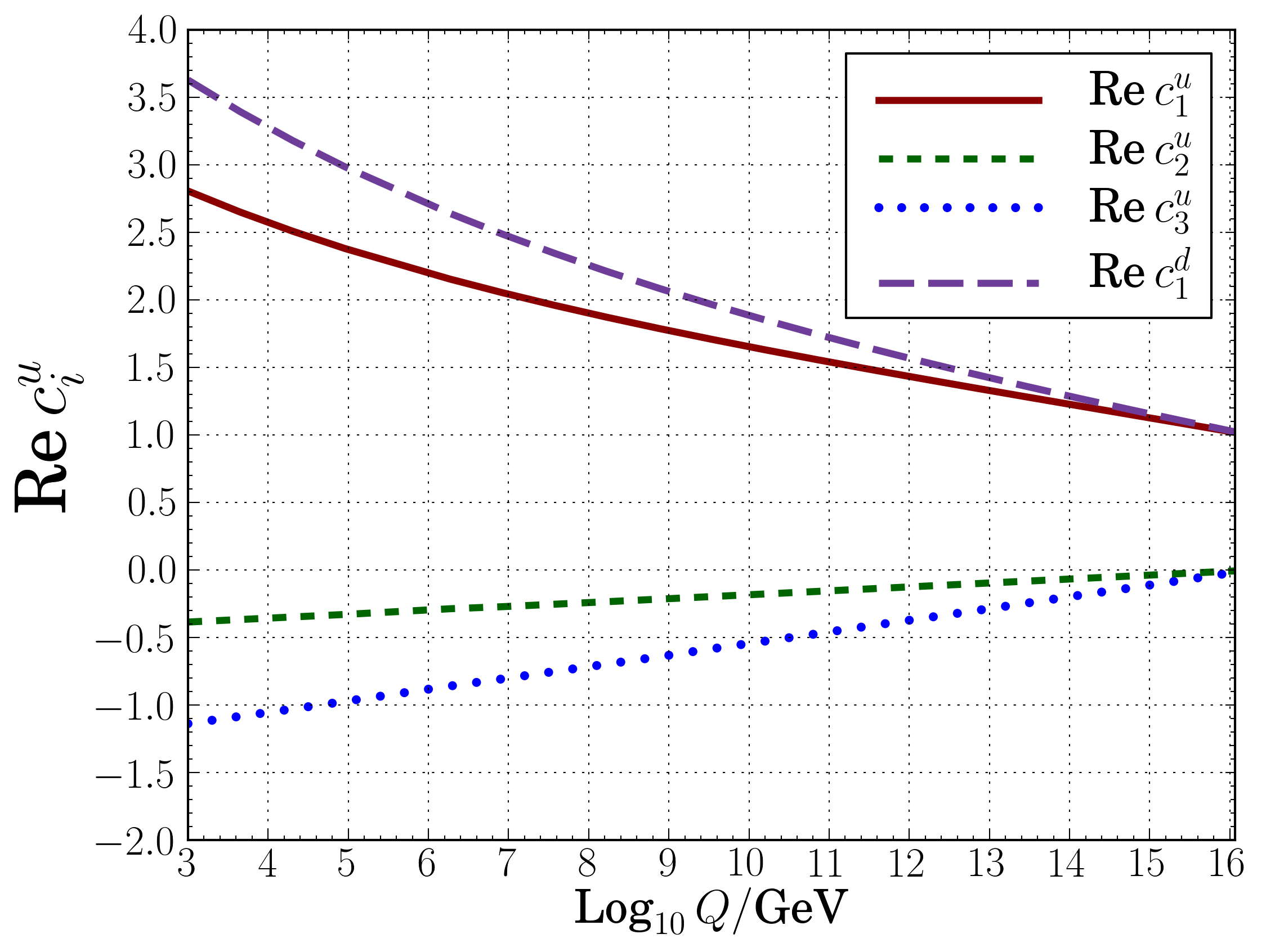}
\caption{Evolution of the leading expansion coefficients for the strictly 
MFV ``natural SUSY'' scenario with light $\tilde t_{L,R}$ and 
$\tilde b_L$ but a heavy $\tilde b_R$. Concretely, we take   
$m_0=10$~TeV, $m_{1/2}=1$~TeV, $A_0=-1$~TeV, $\tan\beta=10$, 
$m_{H_u}^2=m_{H_d}^2=7.5$~(TeV)$^2$, and $\alpha_q=\alpha_u=0.97$. 
The resulting spectrum has $m_{\tilde t_1}=555$~GeV, $m_{\tilde b_1}=570$~GeV, $m_{\tilde t_2}\simeq 1.8$~TeV and all other squark masses $\approx 10$~TeV; moreover $\mu\simeq 800$~GeV and $m_{\tilde g}\simeq 2.5$~TeV. 
The point has a light Higgs mass of $m_h=124$~GeV and passes flavour constraints 
(computed with {\tt SUSY\_FLAVOR 2.02}~\cite{SusyFlavor}). 
Finally, $m_A\simeq 3$~TeV, so we are deep in the Higgs decoupling regime.
\label{fig:heavysbR}}
\end{figure}

\begin{figure}[t]\centering
\includegraphics[width=0.44\textwidth,clip]{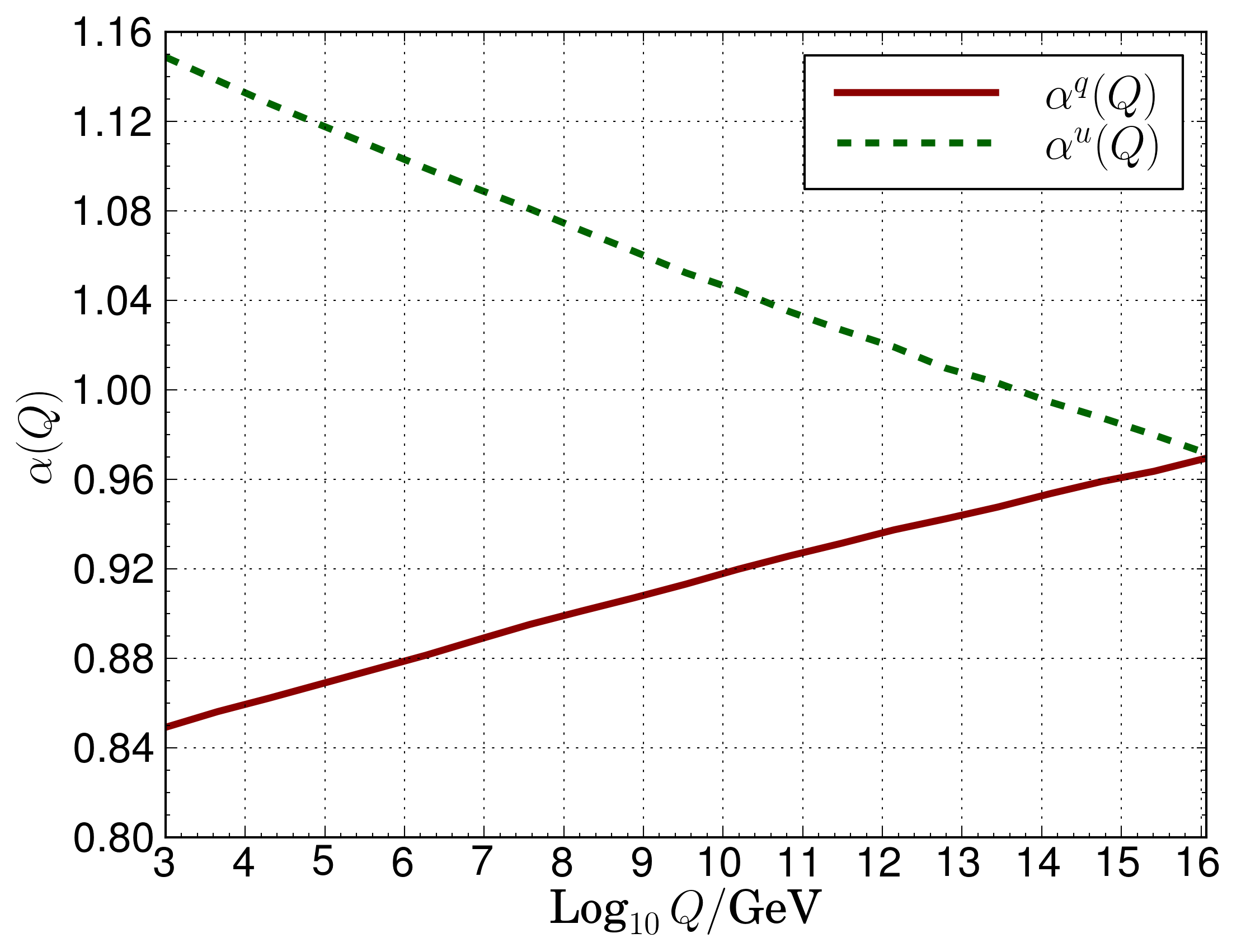}
\caption{Evolution of $\alpha^q$ and $\alpha^u$ within the scenario of Fig.~\ref{fig:heavysbR}. 
The plot serves to confirm the flatness of the evolution of $\alpha$. Moreover, it illustrates that the evolution of the flavour coefficients $a_3^q$ and $a_2^u$ is dominated by the RG evolution of $y_t$, which is factored out here.}
\label{fig:alpha}
\end{figure}

Other scenarios can be constructed along
the same lines. For instance, to also split the $\tilde t_R$ from the first- 
and second-generation squarks, one can further impose%
\begin{equation}\label{alphaqu}
\mathbf{m}_{U}^{2}=m_{0}^{2}\bigl( \mathbf{1}-\alpha_{u}\mathbf{Y}%
_{u}^{\dagger}\mathbf{Y}_{u}\langle\mathbf{Y}_{u}^{\dagger}\mathbf{Y}%
_{u}\rangle^{-1}\bigr) \;,
\end{equation}
which is also compatible with the MFV principle when $\alpha_{u}\approx1$.
As opposed to the above scenario, the condition that both $\mathbf{m}_U^2$
and $\mathbf{m}_Q^2$ be hierarchical is radiatively stable (provided that 
the other states which couple strongly to the stop sector, such as the
up-type Higgs and the gauginos, are not too heavy). Together with a
small $\mu$ parameter, this constitutes a way to realize ``natural 
supersymmetry'' within MFV. An example for the 
typical evolution of the leading expansion coefficients 
for such a natural SUSY-MFV scenario is given in Fig~\ref{fig:heavysbR}. 
The RG evolution and computation of the mass spectrum is done with 
{\tt SPheno}~\cite{Porod:2003um,Porod:2011nf} with boundary conditions adapted 
according to Eqs.~(\ref{Expand1})--(\ref{Expand3}). 
The $a_1$ coefficients are not shown because they remain very close to unity,
with deviations at the level of less than a percent. 
The evolution of the $a_3^q$ and $a_2^u$ coefficients is much steeper than that 
of the other $a_i$. The reason for this is that $a_3^q$ and $a_2^u$ are dominated by the running 
of $y_t$; when the $y_t$ dependence is factored out, the evolution is very flat, see Fig.~\ref{fig:alpha}.

\begin{figure}[t!]\centering
\includegraphics[width=0.44\textwidth,clip]{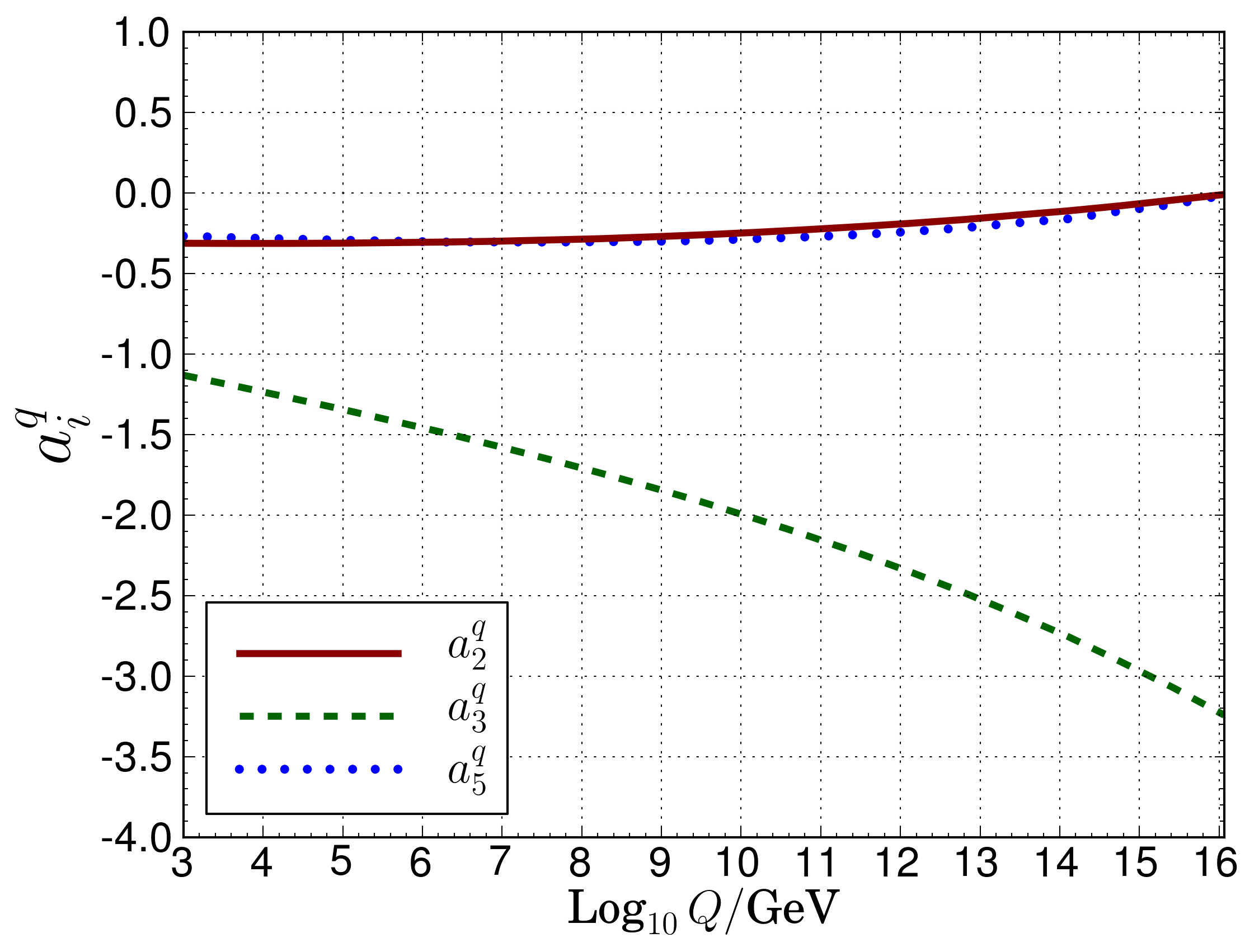}\qquad
\includegraphics[width=0.44\textwidth,clip]{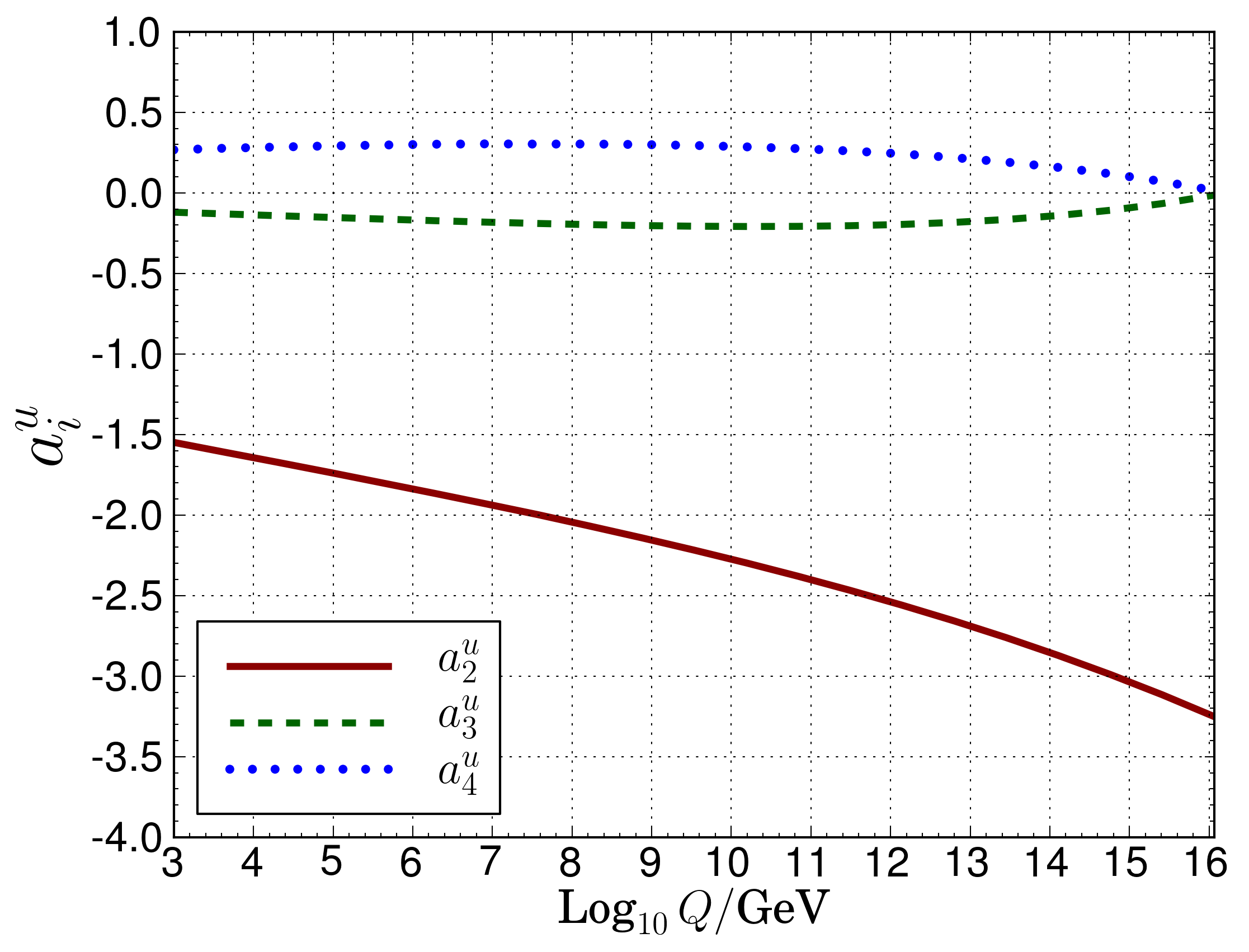}\\
\includegraphics[width=0.44\textwidth,clip]{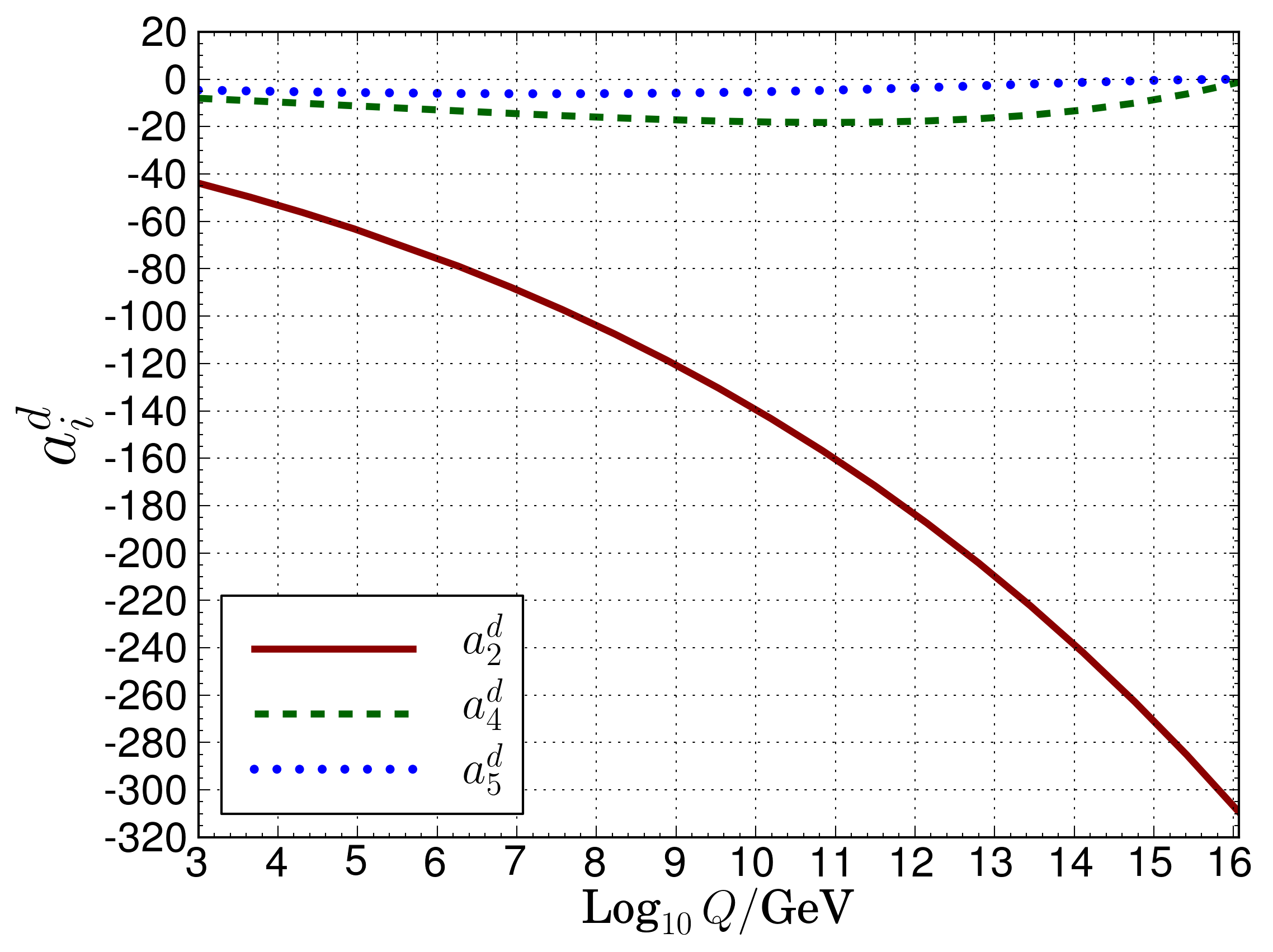}\qquad
\includegraphics[width=0.44\textwidth,clip]{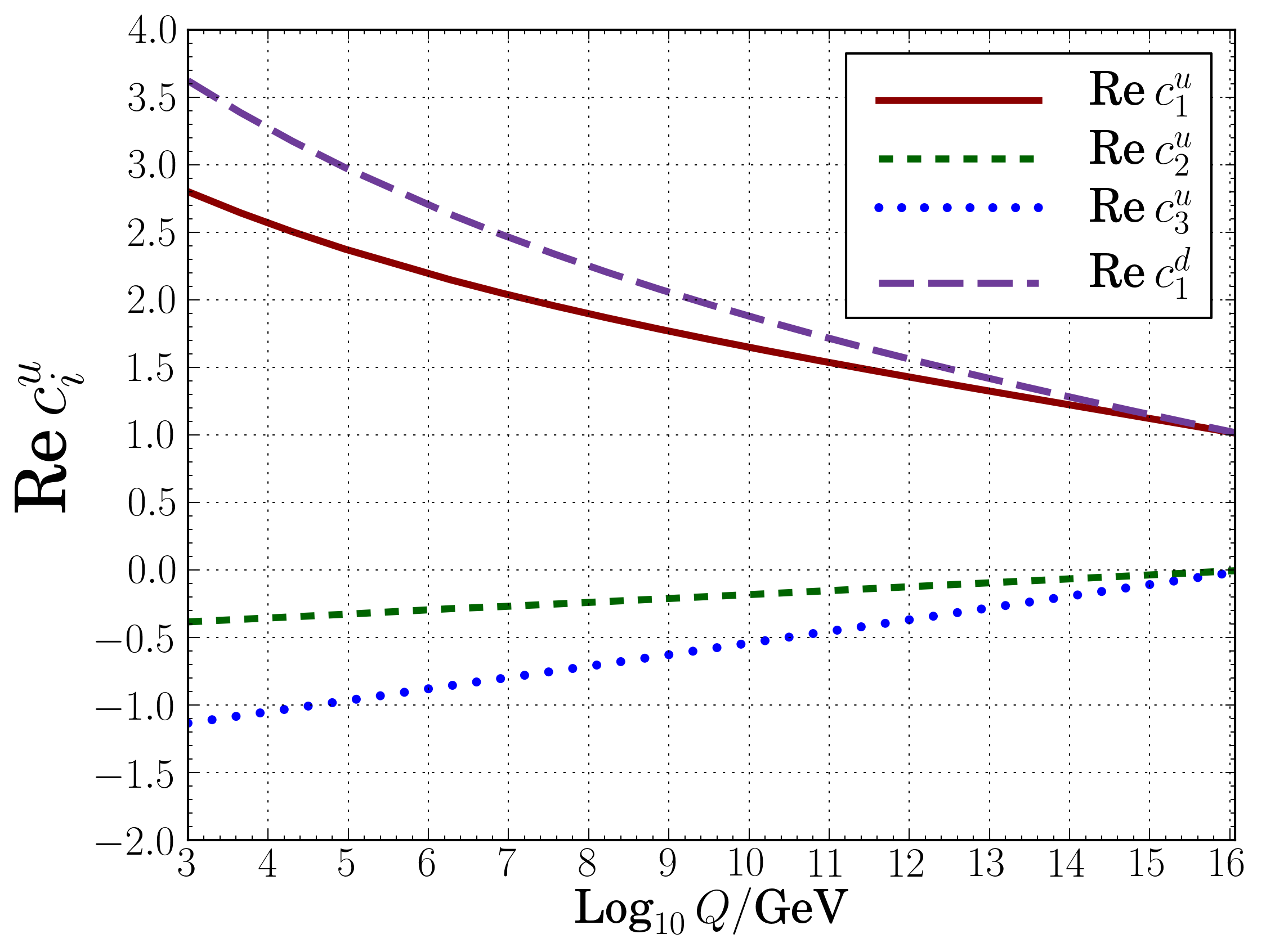}
\caption{Evolution of the leading expansion coefficients in scenario 2, which is not quite MFV 
because the $\tilde b_R$ is also light. Here, we take  
$\alpha_d=315$. The other parameters are as in Fig.~\ref{fig:heavysbR}, apart from adjusting 
$m_{H_u}^2=m_{H_d}^2=5$~(TeV)$^2$ to obtain a $m_h$ near 125~GeV. 
The resulting spectrum is $m_{\tilde t_1}=796$~GeV, $m_{\tilde b_1}\simeq m_{\tilde t_2}\simeq 1.4$~TeV, and 
$m_{\tilde b_2}\simeq 2.4$~TeV. The first/second generation squark masses are again $\approx 10$~TeV. 
The higgsino mass turns quite low, $\mu\simeq 240$~GeV, while the gluino and the additional Higgs states 
remain heavy, $m_{\tilde g}\simeq 2.4$~TeV and $m_A\simeq 2$~TeV.
\label{fig:lightsbR}}
\end{figure}

On the other hand, there is no way to split the right sbottom from the first
two generations without moving away from MFV. Indeed, all the non-trivial
terms in the expansion of $\mathbf{m}_{D}^{2}$ are sandwiched between
$\mathbf{Y}_{d}^{\dagger}$ and $\mathbf{Y}_{d}$, which are small when
$\tan\beta$ is not very large. Specifically, the simplest way to lighten all
third generation squarks is to impose%
\begin{align}
\mathbf{m}_{Q}^{2} &  =m_{0}^{2}\bigl( \mathbf{1}-\alpha_{q}\mathbf{Y}%
_{u}\mathbf{Y}_{u}^{\dagger}\langle\mathbf{Y}_{u}\mathbf{Y}_{u}^{\dagger
}\rangle^{-1}\bigr) ^{\mathrm{T}}\;,\nonumber\\
\mathbf{m}_{U}^{2} &  =m_{0}^{2}\bigl( \mathbf{1}-\alpha_{u}\mathbf{Y}%
_{u}^{\dagger}\mathbf{Y}_{u}\langle\mathbf{Y}_{u}^{\dagger}\mathbf{Y}%
_{u}\rangle^{-1}\bigr) \;,\nonumber\\
\mathbf{m}_{D}^{2} &  =m_{0}^{2}\bigl( \mathbf{1}-\alpha_{d}\mathbf{Y}%
_{d}^{\dagger}\mathbf{Y}_{d}\langle\mathbf{Y}_{d}^{\dagger}\mathbf{Y}%
_{d}\rangle^{-1}\bigr) \;,\\
\mathbf{T}_{u,d} &  =A_{0}\mathbf{Y}_{u,d}\;, \nonumber
\end{align}
with $\alpha_{q,u,d}\approx1$. Clearly, unless $\tan\beta$ is very large,
$\mathbf{m}_{D}^{2}$ significantly deviates from the MFV assumption. 
One might worry that this setting conflicts with current flavour constraints, 
which would thus disfavour light $\tilde b_R$ squarks. 
However, this is not the case. First, note that a large $a_{2}^{d}\sim\langle\mathbf{Y}_{d}^{\dagger
}\mathbf{Y}_{d}\rangle^{-1}\approx y_{b}^{-2}$ at the low scale is harmless, since it does not contribute to the $\delta_{RR}^{d}$ mass insertions (this is evident in a basis where $\mathbf{Y}_{d}$ is diagonal). The impact of a large $a_{2}^{d}$ at the high scale is less obvious, since it can drive other coefficients towards large non-MFV values through the RGE evolution. However, as illustrated in Fig.~\ref{fig:lightsbR}, this effect turns out to be quite limited numerically. Though some coefficients are indeed initially driven towards large values, the quasi-fixed point behaviour of the RGE evolution then kicks in and brings them back to MFV-like values at the low scale (see {\it e.g.} the coefficient $a_4^d$ in Fig.~\ref{fig:lightsbR}). So, even if the low-scale coefficients are not strictly compatible with the MFV principle, they are sufficiently close to MFV to pass all flavour constraints (we also checked this explicitly by direct computation of the flavour observables, using the {\tt SUSY\_FLAVOR 2.02}
code~\cite{SusyFlavor}).

There is another scenario worth considering. Imagine that for some
reasons, the shift from universality induced by the yet unknown flavour
dynamics occurs only in the $\mathrm{SU}(3)_{Q}$ space, through the
$\mathbf{Y}_{u}\mathbf{Y}_{u}^{\dagger}-\langle\mathbf{Y}_{u}\mathbf{Y}%
_{u}^{\dagger}\rangle$ combination. Plugging this structure in the
soft-breaking expansion, they can be parametrized at the scale
$M_{\mathrm{GUT}}$:
\begin{align}
\mathbf{m}_{Q}^{2} &  =m_{0}^{2}a_{1}^{q}\bigl( \mathbf{1}-\alpha
_{0}\mathbf{Y}_{u}\mathbf{Y}_{u}^{\dagger}\langle\mathbf{Y}_{u}\mathbf{Y}%
_{u}^{\dagger}\rangle^{-1}\bigr) ^{\mathrm{T}}\;,\nonumber \\
\mathbf{m}_{U}^{2} &  =m_{0}^{2}\bigl( a_{1}^{u}\mathbf{1}+a_{2}^{u}%
\mathbf{Y}_{u}^{\dag}(\mathbf{1}-\alpha_{0}\mathbf{Y}_{u}\mathbf{Y}%
_{u}^{\dagger}\langle\mathbf{Y}_{u}\mathbf{Y}_{u}^{\dagger}\rangle
^{-1})\mathbf{Y}_{u}\bigr) \approx m_{0}^{2}a_{1}^{u}\,\mathbf{1\;,}\nonumber \\
\mathbf{m}_{D}^{2} &  =m_{0}^{2}\bigl( a_{1}^{d}\mathbf{1}+a_{2}^{d}%
\mathbf{Y}_{d}^{\dag}(\mathbf{1}-\alpha_{0}\mathbf{Y}_{u}\mathbf{Y}%
_{u}^{\dagger}\langle\mathbf{Y}_{u}\mathbf{Y}_{u}^{\dagger}\rangle
^{-1})\mathbf{Y}_{d}\bigr) \approx m_{0}^{2}a_{1}^{d}\,\mathbf{1}\;,\; \\
\mathbf{T}_{u,d} &  =c_{1}^{u,d}A_{0}\mathbf{Y}_{u,d}\bigl( \mathbf{1}%
-\alpha_{0}\mathbf{Y}_{u}^{\dagger}\mathbf{Y}_{u}\langle\mathbf{Y}%
_{u}\mathbf{Y}_{u}^{\dagger}\rangle^{-1}\bigr) \;, \nonumber
\end{align}
for some $\mathcal{O}(1)$ coefficients $a_{i}^{q,u,d}$ and $c_{i}^{u,d}$,
which we set to one for simplicity. Note how the $a_{2}^{u,d}$ terms end up
negligible because $\mathbf{1}-\alpha_{0}\mathbf{Y}_{u}\mathbf{Y}_{u}%
^{\dagger}\langle\mathbf{Y}_{u}\mathbf{Y}_{u}^{\dagger}\rangle^{-1}$, whose
$(3,3)$ entry is suppressed, is sandwiched between $\mathbf{Y}_{u,d}^{\dag}$
and $\mathbf{Y}_{u,d}$. Again, this input respects the MFV requirement. The
only difference with the first scenario is to impose inverted hierarchies in
the trilinear terms at the unification scale. 
Such a pattern does not survive to the evolution however. Looking at the expansion of the trilinear terms, the leading $c_1^{u,d}$ and subleading $c_{i\neq1}^{u,d}$ coefficients do not evolve at the same speed, especially when the former are driven by the gluino mass. So, the cancellation present at the unification scale does not happen at the low scale, and trilinear terms end up being quite similar to those obtained with the first scenario. In this respect, the difficulty mentioned there to obtain a viable spectrum applies here also; a dedicated numerical analysis would be needed to conclude on the valid parameter space of these scenarios. 

Beyond these specific examples, it is now straightforward to state a more general
sufficient condition for obtaining a GUT-scale split spectrum which is guaranteed to be
flavour-safe, using our formalism. This condition is that the GUT-scale flavour coefficients 
should at most be ${\cal O}(1)$ and should approximately satisfy the relations
(generalizing the expressions for $\mathbf{m}_Q^2$ in Eq.~\eqref{eq:scenario1} and 
$\mathbf{m}_U^2$ in Eq.~\eqref{alphaqu})
\be\begin{split}\label{sumrule}
a_1^q+a_3^q\,y_t^{-2}+a_5^q\,y_t^{-4}&=0\,,\\
a_1^u+a_2^u\,y_t^{-2}+a_4^u\,y_t^{-4}&=0\,.
\end{split}
\ee
The MFV condition ensures that there are no flavour problems, while the sum rules Eqs.~\eqref{sumrule}
ensure that the top squarks are actually split from the first two generation up-type squarks (note
that only $a_1^q$, $a_3^q$ and $a_5^q$ can significantly contribute to the LH
stop soft mass if all $a_i^q$ are $\lesssim{\cal O}(1)$, and similarly for $a_1^u$, $a_3^u$ and $a_5^u$
and the RH stop mass). 

While this prescription covers a large class of viable spectra, we note that it is of course
also possible to obtain flavour safe natural SUSY mass patterns in a different manner
--- for instance, as we have seen above, one may deviate from the MFV prescription by splitting
also the right-handed sbottom mass, and rely on the RG evolution to produce an almost MFV spectrum
at the low scale. For such scenarios, however, safeness from FCNC constraints is not automatic but must
be checked in each case.

We also note that the above sum rules are tied to small or moderately large $\tan\beta$.
At very  large $\tan\beta$, where $y_b$ is of order one, they should be modified to take into account
also the remaining terms in Eqs.~\eqref{Expand1} and \eqref{Expand2}, which may now contribute to
the third-generation squark masses even if their coefficients are ${\cal O}(1)$.

\section{Conclusions}

Third-generation squarks below the TeV scale are an essential requirement for
supersymmetry to be natural, while the squarks of the first two generations
are likely much heavier. Therefore it is important to study the physics of
non-universal squark masses, and of inverted squark mass hierarchies in
particular. In phenomenological approaches which prescribe the soft terms 
at the TeV scale, such as the pMSSM, this is possible to a limited
extent only, since effects arising from the renormalization group running from the
mediation scale are not accounted for. In particular, these effects could
lead to radiatively induced flavour-violating squark mass mixings. Given the tight experimental constraints from flavour observables, to fully grasp the implications of non-universal squark masses, one should be careful to account for such effects.

In this paper we have studied non-universal squark masses in the case 
that SUSY breaking is mediated at the GUT scale. We have shown how split squark 
mass matrices (and trilinears) can be conveniently and generally prescribed in 
a basis-independent way, and investigated their renormalization group evolution.

When requiring only the top squarks to be light, and the first two generations
to be nearly mass degenerate, the most natural prescription automatically respects 
the principle of minimal flavour violation at the GUT scale. Since MFV is preserved during the RG evolution of the soft terms down to the TeV scale, bounds on FCNCs can be easily evaded. 

For more general hierarchical soft terms at the GUT scale, the compatibility with flavour observables is not automatic, even though generic soft terms tend to be attracted towards MFV-like structures in the infrared~\cite{Paradisi:2008qh,Colangelo:2008qp}. We have confirmed this tendency for the particularly relevant case where all third-generation squarks, including the right-sbottom, are light compared to the squarks of the first two generations. While this scenario strongly violates the MFV hypothesis at the GUT scale, the soft terms become increasingly MFV-like during the running, and end up compatible with flavour constraints at the low scale.

Our analysis puts the increasingly popular framework of ``natural SUSY'' on
a more solid footing, showing that it is actually possible to obtain a
natural SUSY spectrum at the TeV scale from well-motivated
GUT-scale boundary conditions without having to worry about RG-induced
flavour violation. Furthermore, our formalism for defining non-universal
soft terms in a basis independent way should be very useful for further
studies of the supersymmetric flavour problem beyond minimal flavour
violation. 
A full exploration, within our scheme, of the parameter space leading to natural SUSY is left for a subsequent work.

\section*{Acknowledgements} 

We thank B.~Allanach, W.~Porod and F.~Staub for clarifying discussions on the 
treatment of flavour violation in {\tt SOFTSUSY} and {\tt SPheno}.  
Moreover, we specifically thank W.~Porod for help with adapting {\tt SPheno} to our needs.

This work originated from the workshop ``Implications of the 125~GeV Higgs boson'', 
which was held 18--22 March 2013 at LPSC Grenoble and which was partially funded 
by the LabEx ENIGMASS and the Centre de Physique Th\'eorique de Grenoble (CPTG).

\section*{Appendix: Stability of the expansion coefficients}

\addtocounter{section}{1}

The CKM matrix plays a central role in the description of flavour mixing in the quark and squark sectors. Two numerical approximations are often introduced: the CP-conserving limit and the neglect of threshold corrections. At first sight, it may appear reasonable to use an approximate CKM matrix in the running to and from the unification scale. After all, the error should be small, and one can always plug back the exact CKM matrix for computing flavour observables. However, while this procedure obviously suffices to bring back the quark mixing to its physical value, this is not the case in the squark sector. Indeed, in many scenarios, the off-diagonal entries in the squark soft terms at the electroweak scale are entirely driven through RG running from the CKM matrix. For example, starting with universal boundary conditions, flavour mixing in the left-squark soft mass term is given by
\begin{equation}
\mathbf{M}_{\tilde d}^{LL}[1\;\text{TeV}]^{I\neq J}\sim(\mathbf{Y}_{u}\mathbf{Y}_{u}^\dag)^{JI}\sim y_{t}^{2}V_{tI}^{\ast}V_{tJ}\;,
\label{SoftYuYu}
\end{equation}
since $v_{u}\mathbf{Y}_{u}^{\mathrm{T}}=\mathbf{M}_{u}\cdot\mathbf{V}_{\mathrm{CKM}}$ in the down-quark mass eigenstate basis (in Eq.~(\ref{SoftYuYu}), CKM entries are conventionally denoted as $V_{IJ}$, with $I=u,c,t$ and $J=d,s,b$ instead of $I,J=1,2,3$). Therefore, if a wrong CKM matrix is used throughout the running, the soft terms are also wrong, and so are the estimated supersymmetric contributions to the FCNC processes.

In the present section, our goal is to show that these issues can be circumvented if the squark soft mass terms and trilinear terms are defined through their expansion coefficients. Indeed, to a large extent, these do not depend on the precise value of the CKM matrix entries. So, once the expansion coefficients at the low scale have been computed under some approximation, it is a simple matter to reconstruct with an excellent accuracy the physical soft terms by plugging back the physical CKM matrix. Let us illustrate this procedure.

\subsection*{CP-conserving limit for the CKM matrix}

As a first approximation, the MSSM evolution is often computed in the CP-conserving limit. To this end, the CP violating phase of the SM must somehow be disposed of. There is no unique way to achieve this, since there is no unique way to parametrize the CKM matrix itself, and no matter the chosen
procedure, the modulus of at least one of the CKM entries is significantly affected.

Let us take ${\bf m}_Q^2$ as an example. If the true, complex CKM matrix is used, and using the same scenario as in Fig.~\ref{fig:heavysbR}, then 
\begin{align}
\frac{{\bf m}_Q^2[1\;\text{TeV}]}{m_{0}^{2}}&=\left(
\begin{array}[c]{ccc}%
0.9932 & 0.3395\times10^{-3} & -0.8146\times10^{-2}\\
0.3395\times10^{-3} & 0.9916 & 0.4085\times10^{-1}\\
-0.8146\times10^{-2} & 0.4085\times10^{-3} & 0.4543\times10^{-2}
\end{array}\right) \nonumber\\[2mm]
&\hspace{1cm}+\,i\left(\begin{array}[c]{ccc}%
0 & 0.1414\times10^{-3} & -0.3250\times10^{-2}\\
-0.1414\times10^{-3} & 0 & -0.0075\times10^{-1}\\
0.3250\times10^{-2} & 0.0075\times10^{-3} & 0
\end{array}\right)  \,.
\label{mdLLexact}
\end{align}
These numbers are obtained starting with the Wolfenstein parameters \cite{Charles:2004jd} $\lambda =0.22457, A = 0.823,\bar{\rho}=0.1289,\bar{\eta}=0.348$, using SM RGEs up to one TeV, and then MSSM RGEs between 1~TeV and $M_{\rm GUT}$. Threshold corrections for the quark masses and the gauge couplings are taken from {\tt SPheno}~\cite{Porod:2003um,Porod:2011nf}. Projecting ${\bf m}_Q^2[1\;\text{TeV}]$, the expansion coefficients at that scale are
\begin{align}
a_{1\ldots 6} &  =(0.9933,\;-0.3297,\;-1.1499,\;-0.01476,\;-0.2474,\;0.001419),\;\;\\
b_{1\ldots 3} &  =(3.695,\;-4.989,\;0.5002)\times10^{-2}\;.
\end{align}
The purely CP-violating coefficients $b_{1\ldots 3}$ are entirely induced through the RG running. Numerically, their contributions to $\Im({\bf m}_Q^2)$ are extremely suppressed because they are  tuned by the small Jarlskog invariant $\operatorname{Im}\langle(\mathbf{Y}_{u}\mathbf{Y}_{u}^{\dagger})^{2}\mathbf{Y}_{d}\mathbf{Y}_{d}^{\dagger}\mathbf{Y}_{u}\mathbf{Y}_{u}^{\dagger}(\mathbf{Y}_{d}\mathbf{Y}_{d}^{\dagger})^{2}\rangle\sim10^{-5}$. The bulk of $\Im({\bf m}_Q^2)$ actually comes from $\Im(\mathbf{Y}_{u}\mathbf{Y}_{u}^\dag)$, see Eq.~(\ref{SoftYuYu}).

Let us now compare this with the results in the CP-conserving limit. The most frequent CP-conserving prescription is to set $\delta_{13}=0$ in the conventional CKM parametrization. This is the prescription adopted in the RGE codes {\tt SPheno}~\cite{Porod:2011nf} and {\tt SOFTSUSY}~\cite{Allanach:2001kg}.\footnote{Reference~\cite{Allanach:2001kg} explicitly mentions that the CP-conserving limit may induce significant uncertainties.} 
The only CKM entry significantly affected by this is $V_{td}$, 
\begin{equation}
|V_{td}|^{\delta_{13}=0}=0.0058\;\;\;vs.\;\;|V_{td}|=0.0085\;.
\end{equation}
As a consequence of Eq.~(\ref{SoftYuYu}), the $(1,2)$ and $(1,3)$ entries of ${\bf m}_Q^2$ are then significantly reduced, since they are induced by $V_{tb}^{\ast}V_{td}$ and $V_{ts}^{\ast}V_{td}$ respectively:
\begin{equation}
\frac{{\bf m}_Q^2[1\;\text{TeV}]^{\delta_{13}=0}}{m_{0}^{2}}=\left(
\begin{array}
[c]{ccc}
0.9933 & 0.2510\times10^{-3} & -0.5884\times10^{-2}\\
0.2510\times10^{-3} & 0.9916 & 0.4137\times10^{-1}\\
-0.5884\times10^{-2} & 0.4137\times10^{-1} & 0.4543\times10^{-4}
\end{array}
\right) .
\end{equation}
If used to compute FCNC observables, this approximation is particularly dangerous for the $b\rightarrow d$ and $s\rightarrow d$ transitions.\ First, the SM and charged Higgs contributions to $Z,\gamma$ penguins and boxes, dominated by the top quark contributions hence tuned by $V_{tb}^{\ast}V_{td}$, are systematically underestimated. This could still be cured by plugging back the correct CKM matrix in the relevant vertices. This procedure fails, however, to cure the also underestimated gaugino-induced FCNC contributions tuned by $(\mathbf{M}_{\tilde d}^{LL})^{13}$ and $(\mathbf{M}_{\tilde d}^{LL})^{12}$. 

On the other hand, it is easy to check that the expansion coefficients discussed above stay very close to the ones obtained in the CP-violating case. If we project ${\bf m}_Q^2[1\;\text{TeV}]^{\delta_{13}=0}$ using the CP-conserving Yukawa matrices ${\bf Y}_{u,d}[1\;\text{TeV}]^{\delta_{13}=0}$, we find
\begin{align}
a_{1\ldots 6}^{\delta_{13}=0} &  =(0.9933,\;-0.3301,\;-1.1502,\;-0.01445,\;-0.2468,\;0.001130),\;\;\\
b_{1\ldots 3}^{\delta_{13}=0} &  =(0,\;0,\;0)\;.
\end{align}
This remains true for all the other soft terms: the shift in the coefficients is below the percent level for the first five coefficients, and of a few percent for the last four. Thanks to this stability, we can use the coefficients computed in the CP-conserving limit together with the true, CP-violating Yukawa couplings to reconstruct the true CP-violating soft-breaking terms with an excellent accuracy. To be precise, this means that if we compute
\begin{equation}
\left({\tilde {\bf m}}_Q^2\right)^\trans =m_0^2\Bigl(a_1^{\delta_{13}=0}\,{\bf 1}+a_2^{\delta_{13}=0}\,\mathbf{Y}_{u}\mathbf{Y}_{u}^{\dag}+a_3^{\delta_{13}=0}\,\mathbf{Y}_{d}\mathbf{Y}_{d}^{\dag}+...\Bigr)\;, 
\end{equation}
where $\mathbf{Y}_{d} = {\bf Y}_{d}^{\delta_{13}=0}$ but $\mathbf{Y}_{u}^{\mathrm{T}} = ({\bf Y}_{u}^{\mathrm{T}})^{\delta_{13}=0}\cdot \mathbf{V}_{\mathrm{CKM}}^{\delta_{13}=0\; {\dag}}\cdot \mathbf{V}_{\mathrm{CKM}}$, then $|{\tilde {\bf m}}_Q^2 - {\bf m}_Q^2| < 10^{-9}\times m_0^2$. Since such small differences are irrelevant phenomenologically, and since the other soft-breaking terms are equally well reproduced, it is a simple matter to cure at the same time all the contributions to the FCNC from the artefacts of the CP-conserving limit. In practice, it is thus possible to perform the RGE study in the CP-conserving limit, use our prescription on the output file to reconstruct the full-fledged CP-violating flavour structures, and then pass it on to codes like {\tt SUSY\_FLAVOR}~\cite{SusyFlavor} to compute reliably the supersymmetric contributions to the FCNC. This is what we actually did to check the compatibility of the scenarios described in the main text with current flavour constraints.

This procedure works no matter the CP conserving prescription. Let us compare, for instance, 
the $\delta_{13}\rightarrow0$ limit to the $\eta\rightarrow0$ limit. In the latter case, $|V_{td}|$ is reduced only by about 7\%, while $V_{ub}$ is suppressed by nearly 60\%,
\begin{equation}
|V_{ub}|^{\eta=0}=0.00132\;\;\;vs.\;\;|V_{ub}|=0.00349\;.
\end{equation}
However, an underestimated $V_{ub}$ does not bear serious consequences because it does not affect the top sector. Loop level FCNC are insensitive to this reduction since $d^{I}\rightarrow d^{J}$ transitions are
dominantly tuned by $V_{tI}^{\ast}V_{tJ}$. For the same reason, the soft mass terms are closer to the true CP-violating ones, with 
\begin{equation}
\frac{{\bf m}_Q^2[1\;\text{TeV}]^{\eta=0}}{m_{0}^{2}}=\left(
\begin{array}
[c]{ccc}%
0.9932 & 0.3420\times10^{-3} & -0.8147\times10^{-2}\\
0.3420\times10^{-3} & 0.9916 & 0.4085\times10^{-1}\\
-0.8147\times10^{-2} & 0.4085\times10^{-1} & 0.4532\times10^{-4}
\end{array}
\right)  \,,
\end{equation}
which nearly matches the real part of the CP-violating result (but stays significantly off for the absolute parts). This can be understood from Eq.~(\ref{SoftYuYu}): the RGE corrections proportional to $\mathbf{Y}_{u}\mathbf{Y}_{u}^\dag$ depend, to an excellent approximation, only on the third row of the CKM matrix, which stays close to the true one. 
The $\eta\rightarrow0$ limit therefore mostly affects tree-level charged-current flavour-changing observables like $B\rightarrow\tau\nu$, and this is easily cured by plugging back the true value for the CKM matrix.
In any case, the expansion coefficients extracted in the $\eta = 0$ limit are again very close to those obtained in the CP-violating case:
\begin{align}
a_{1\ldots 6}^{\eta =0} &  =(0.9933,\;-0.3299,\;-1.1506,\;-0.01037,\;-0.2463,\;0.000996),\;\;\\
b_{1\ldots 3}^{\eta =0} &  =(0,\;0,\;0)\;.
\end{align}
From them, the reconstructed soft term ${\tilde {\bf m}}_Q^2$ matches ${\bf m}_Q^2$ up to corrections of the order of $10^{-7}\times m_0^2$, which is again more than enough phenomenologically.

It should be stressed here that our prescription works particularly well when the soft-breaking terms respect the MFV hypothesis, {\it i.e.}, when none of the leading expansion coefficients are exceedingly large. In that case, their values are extremely resilient to changes in the CKM parameters, and the prescription reproduces the soft-breaking terms with an impressive accuracy. Beyond MFV, the coefficients in the CP-conserving and violating cases are not necessarily as close. For example, taking the scenario detailed in Fig.~\ref{fig:lightsbR}, we find that coefficients vary by up to about 20\%. But, crucially, these variations affect mostly the subleading coefficients, whose phenomenological impact is very limited. As a result, the CP-conserving coefficients still permit to reconstruct the full CP-violating soft-breaking terms with an accuracy better than 1\%. Thus, even though we have not tested extensively the range of validity of the prescription when moving out of the MFV framework, we expect it 
remains accurate for a broad range of flavour-compatible scenarios.

\subsection*{Threshold corrections and experimental errors on the CKM matrix}

\begin{figure}[t!]\centering
\includegraphics[width=0.44\textwidth,clip]{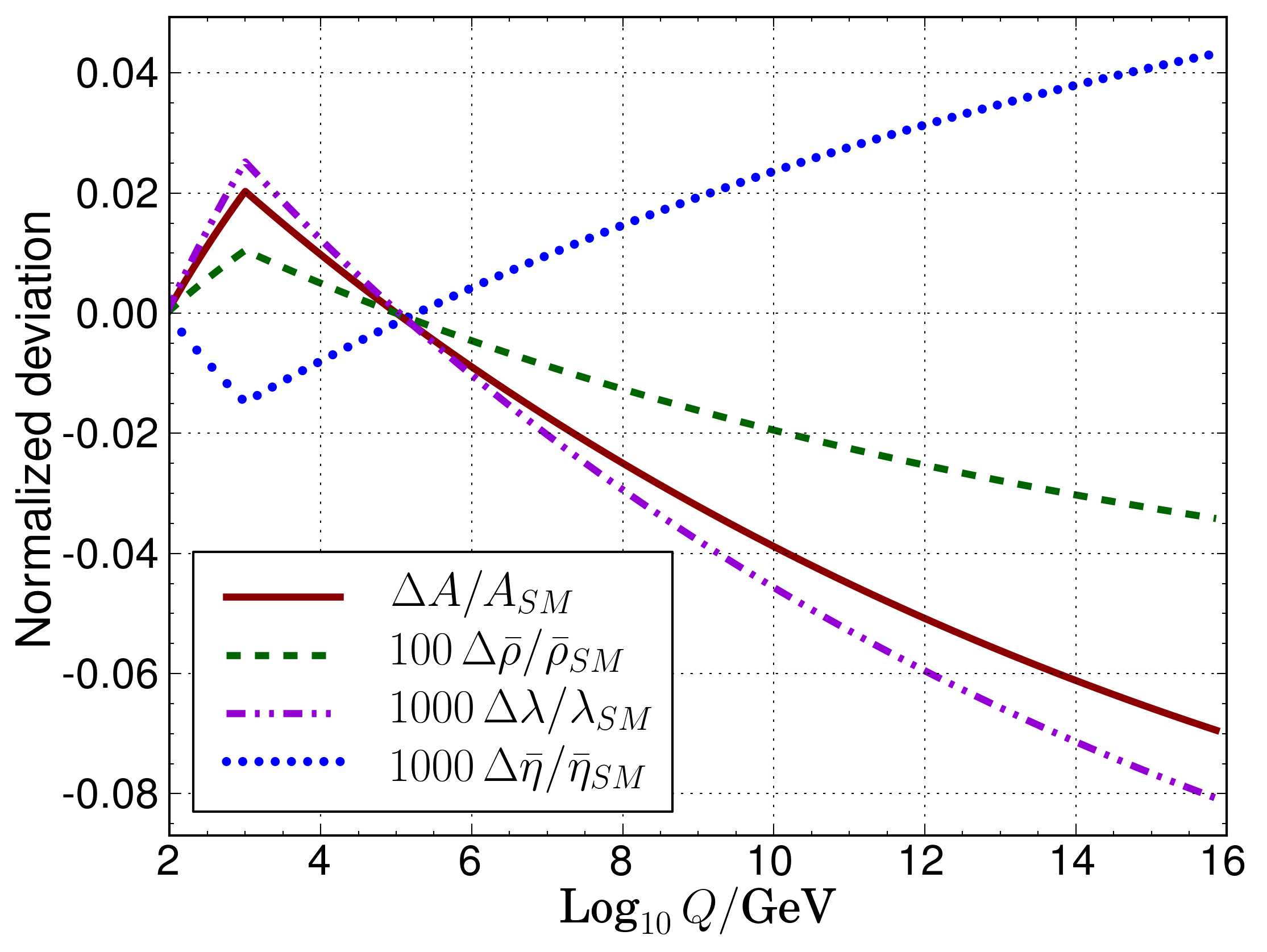}
\caption{Evolution of the Wolfenstein parameters in the scenario of Fig.~\ref{fig:heavysbR}. The evolution is first performed using the SM RGEs up to $1$~TeV, and then according to the MSSM RGEs up to the GUT scale. Plotted are the normalized deviations $\Delta X / X_{SM}, X = \lambda, A, \bar\rho,\bar\eta$, with $\Delta X = X[Q]-X_{SM}$ and $X_{SM} = X[M_Z]$ (the experimental values at $M_Z$ are quoted in Eq.~(\ref{CKMparam})). These deviations are enhanced by 100 (1000) for $\bar\rho$ ($\lambda, \bar\eta$) for better visibility.
\label{fig:CKMparam}}
\end{figure}

It is well known that the CKM matrix runs very slowly. So, for simplicity, when it is evolved using the MSSM beta functions already from the electroweak scale, it is not subsequently corrected for threshold effects. There is, however, a coincidental fact that tends to slightly enhance the error induced by this procedure: the SM and MSSM beta functions for the CKM parameters have opposite signs~\cite{Babu:1987im}. 
As shown in Fig.~\ref{fig:CKMparam}, neglecting the former, the CKM angles are underestimated at all scales. As a result, CKM-driven flavour mixing in the squark soft-breaking terms, {\it i.e.}\ those arising from both the RGE effects and the GUT-scale boundary conditions, are underestimated.

Numerically, the effect on the CKM parameters is small but not entirely negligible. Let us use their SM running between $M_{Z}$ and $1$~TeV as a measure of their sensitivity to threshold corrections, see Fig.~\ref{fig:CKMparam}. The variations of the Wolfenstein parameters are all much smaller than their corresponding experimental errors, except for the $A$ parameter~\cite{Balzereit:1998id}, which increases by about $2\%$ from $M_{Z}$ to $1$~TeV. In view of this, we can estimate the impact of neglecting CKM threshold effects on the soft terms by decreasing the $A$ parameter. As a rough estimate, we send the $A$ parameter to the low end of its $2\sigma$ range, $A = 0.823^{+0.018}_{-0.042}$. 
Still using the scenario of Fig.~\ref{fig:heavysbR}, this leads to
\begin{align}
\frac{{\bf m}_Q^2[1\;\text{TeV}]}{m_{0}^{2}}&=\left(
\begin{array}[c]{ccc}%
0.9932 & 0.3061\times10^{-3} & -0.8324\times10^{-2}\\
0.3061\times10^{-3} & 0.9916 & 0.4174\times10^{-1}\\
-0.8324\times10^{-2} & 0.4174\times10^{-3} & 0.4621\times10^{-2}
\end{array}\right) \nonumber \\[2mm]
&\hspace{1cm}+\,i\left(\begin{array}[c]{ccc}%
0 & 0.1476\times10^{-3} & -0.3321\times10^{-2}\\
-0.1476\times10^{-3} & 0 & -0.0077\times10^{-1}\\
0.3321\times10^{-2} & 0.0077\times10^{-3} & 0
\end{array}\right)  \,,
\label{mdLLth}
\end{align}
which deviates by up to about 10\% from the values in Eq.~(\ref{mdLLexact}). This shows that even supposedly negligible shifts in the CKM parameters can build up sizeable effects in the soft-breaking terms. The expansion coefficients, on the other hand,  are the same up to completely negligible shifts of the order of $10^{-7}$. In other words, these coefficients are essentially independent of the threshold corrections even though soft-breaking terms can deviate significantly. So, whenever the threshold effects for the CKM running are not fully taken care of, one can rely on the same strategy as for the CP-limit, {\it i.e.},  compute the coefficients and then reconstruct accurately the soft-breaking terms by plugging in the physical CKM matrix.

As an interesting corollary, the stability of the coefficients offers a very simple procedure to estimate the impact of the CKM experimental errors on the soft-breaking terms. Only one run is needed with the central values of the CKM parameters to get the expansion coefficients, and once known, it suffices to vary the CKM matrix entering the Yukawa couplings used to reconstruct the soft-breaking terms at the low scale. Let us illustrate this procedure. First, we perform the RGE evolution starting with the electroweak-scale CKM matrix obtained by shifting all the Wolfenstein parameters to the extremes of their $2\sigma$ ranges \cite{Charles:2004jd}:
\begin{equation}
\lambda =0.22457^{+0.00193}_{-0.00021},\; A = 0.823^{+0.025}_{-0.049},\; \bar{\rho}=0.129^{+0.056}_{-0.018}, \;\bar{\eta}=0.348^{+0.025}_{-0.030}\;.
\label{CKMparam}
\end{equation}
We do not take into account the correlations between these parameters. The ranges of values for the soft-breaking term entries are then
\begin{align}
\frac{{\bf m}_Q^2[1\;\text{TeV}]}{m_{0}^{2}}&=\left(
\begin{array}[c]{ccc}%
0.99324^{+0.00002}_{-0.00001} & 0.340^{+0.037}_{-0.053}\times10^{-3} & 
-0.815^{+0.093}_{-0.057}\times10^{-2}\\
0.340^{+0.037}_{-0.053}\times10^{-3} & 0.9916^{+0.0002}_{-0.0001} & 
0.409^{+0.016}_{-0.020}\times10^{-1}\\
-0.815^{+0.093}_{-0.057}\times10^{-2} & 0.409^{+0.016}_{-0.020}\times10^{-3} & 
0.454^{+0.014}_{-0.018}\times10^{-2}
\end{array}\right)\\[2mm]
&\hspace{4mm}+\,i\left(\begin{array}[c]{ccc}%
0 & 0.141^{+0.024}_{-0.026}\times10^{-3} & -0.325^{+0.044}_{-0.040}\times10^{-2}\\
-0.141^{+0.024}_{-0.026}\times10^{-3} & 0 & -0.007\pm0.001\times10^{-1}\\
0.325^{+0.044}_{-0.040}\times10^{-2} & 0.007\pm0.001\times10^{-3} & 0
\end{array}\right)  \,. \nonumber
\label{mdLLrange}
\end{align}
This represents sizeable shifts, up to 30\% (40\%) for the real (imaginary) parts. On the other hand, the expansion coefficients do not change significantly: the first five of each expansion being shifted by less than $10^{-6}$, while the last three of each expansion by less than $10^{-4}$. They are thus essentially constant over the experimental ranges for the CKM parameters. This confirms that once the experimental errors on the CKM matrix at a given scale are known, the full RGE analysis needs to be performed only once to derive those on the soft-breaking terms at that scale. Since this rather indirect but nevertheless significant impact of the errors on the CKM matrix elements is in general neglected, this could greatly improve and simplify the study of their effect on the flavour constraints for a given scenario.



\begin{thebibliography}{10}                                                                                                


\bibitem{Cohen:1996vb}
  A.~G.~Cohen, D.~B.~Kaplan and A.~E.~Nelson,
  ``The More minimal supersymmetric standard model,''
  Phys.\ Lett.\ B {\bf 388} (1996) 588
  [hep-ph/9607394].

\bibitem{Kitano:2006gv}
  R.~Kitano and Y.~Nomura,
  ``Supersymmetry, naturalness, and signatures at the LHC,''
  Phys.\ Rev.\ D {\bf 73} (2006) 095004
  [hep-ph/0602096].

\bibitem{Barbieri:2009ev}
  R.~Barbieri and D.~Pappadopulo,
  ``S-particles at their naturalness limits,''
  JHEP {\bf 0910} (2009) 061
  [arXiv:0906.4546 [hep-ph]].

\bibitem{Papucci:2011wy}
  M.~Papucci, J.~T.~Ruderman and A.~Weiler,
  ``Natural SUSY Endures,''
  JHEP {\bf 1209} (2012) 035
  [arXiv:1110.6926 [hep-ph]].

\bibitem{Baer:2012uy}
  H.~Baer, V.~Barger, P.~Huang and X.~Tata,
  ``Natural Supersymmetry: LHC, dark matter and ILC searches,''
  JHEP {\bf 1205} (2012) 109
  [arXiv:1203.5539 [hep-ph]].

\bibitem{Brummer:2012ns}
  F.~Br\"ummer, S.~Kraml and S.~Kulkarni,
  ``Anatomy of maximal stop mixing in the MSSM,''
  JHEP {\bf 1208} (2012) 089
  [arXiv:1204.5977 [hep-ph]].
  
\bibitem{Badziak:2012rf}
  M.~Badziak, E.~Dudas, M.~Olechowski and S.~Pokorski,
  ``Inverted Sfermion Mass Hierarchy and the Higgs Boson Mass in the MSSM,''
  JHEP {\bf 1207} (2012) 155
  [arXiv:1205.1675 [hep-ph]].

\bibitem{Giudice:2008uk}
  G.~F.~Giudice, M.~Nardecchia and A.~Romanino,
  ``Hierarchical Soft Terms and Flavor Physics,''
  Nucl.\ Phys.\ B {\bf 813} (2009) 156
  [arXiv:0812.3610 [hep-ph]].

\bibitem{Kersten:2012ed}
  J.~Kersten and L.~Velasco-Sevilla,
  ``Flavour constraints on scenarios with two or three heavy squark generations,''
  Eur.\ Phys.\ J.\ C {\bf 73} (2013) 2405
  [arXiv:1207.3016 [hep-ph]].

\bibitem{Mescia:2012fg}
  F.~Mescia and J.~Virto,
  ``Natural SUSY and Kaon Mixing in view of recent results from Lattice QCD,''
  Phys.\ Rev.\ D {\bf 86} (2012) 095004
  [arXiv:1208.0534 [hep-ph]].

\bibitem{Barbieri:2010ar}
  R.~Barbieri, E.~Bertuzzo, M.~Farina, P.~Lodone and D.~Zhuridov,
  ``Minimal Flavour Violation with hierarchical squark masses,''
  JHEP {\bf 1012} (2010) 070
   [Erratum-ibid.\  {\bf 1102} (2011) 044]
  [arXiv:1011.0730 [hep-ph]].

\bibitem{Gherghetta:2011wc}
  T.~Gherghetta, B.~von Harling and N.~Setzer,
  ``A natural little hierarchy for RS from accidental SUSY,''
  JHEP {\bf 1107} (2011) 011
  [arXiv:1104.3171 [hep-ph]].

\bibitem{Auzzi:2011eu}
  R.~Auzzi, A.~Giveon and S.~B.~Gudnason,
  ``Flavor of quiver-like realizations of effective supersymmetry,''
  JHEP {\bf 1202} (2012) 069
  [arXiv:1112.6261 [hep-ph]].

\bibitem{Larsen:2012rq}
  G.~Larsen, Y.~Nomura and H.~L.~L.~Roberts,
  ``Supersymmetry with Light Stops,''
  JHEP {\bf 1206} (2012) 032
  [arXiv:1202.6339 [hep-ph]].

\bibitem{Craig:2012di}
  N.~Craig, M.~McCullough and J.~Thaler,
  ``Flavor Mediation Delivers Natural SUSY,''
  JHEP {\bf 1206} (2012) 046
  [arXiv:1203.1622 [hep-ph]].

\bibitem{Eliaz:2013aaa}
  L.~Eliaz, A.~Giveon, S.~B.~Gudnason and E.~Tsuk,
  ``Mild-split SUSY with flavor,''
  JHEP {\bf 1310} (2013) 136
  [arXiv:1306.2956 [hep-ph]].

\bibitem{Dudas:2013pja}
  E.~Dudas, G.~von Gersdorff, S.~Pokorski and R.~Ziegler,
  ``Linking Natural Supersymmetry to Flavour Physics,''
  JHEP {\bf 1401} (2014) 117
  [arXiv:1308.1090 [hep-ph]].

\bibitem{Arvanitaki:2013yja}
  A.~Arvanitaki, M.~Baryakhtar, X.~Huang, K.~Van Tilburg and G.~Villadoro,
  ``The Last Vestiges of Naturalness,''
  arXiv:1309.3568 [hep-ph].

\bibitem{Brummer:2013upa}
  F.~Br\"ummer, M.~McGarrie and A.~Weiler,
  ``Light third-generation squarks from flavour gauge messengers,''
  arXiv:1312.0935 [hep-ph].

\bibitem{Dudas:2013gga}
  E.~Dudas, M.~Goodsell, L.~Heurtier and P.~Tziveloglou,
  ``Flavour models with Dirac and fake gluinos,''
  arXiv:1312.2011 [hep-ph].

\bibitem{D'Ambrosio:2002ex}
  G.~D'Ambrosio, G.~F.~Giudice, G.~Isidori and A.~Strumia,
  ``Minimal flavor violation: An Effective field theory approach,''
  Nucl.\ Phys.\ B {\bf 645} (2002) 155
  [hep-ph/0207036]. 

\bibitem{Allanach:2008qq}
  B.~C.~Allanach, C.~Balazs, G.~Belanger, M.~Bernhardt, F.~Boudjema, D.~Choudhury, K.~Desch and U.~Ellwanger {\it et al.},
  ``SUSY Les Houches Accord 2,''
  Comput.\ Phys.\ Commun.\  {\bf 180} (2009) 8
  [arXiv:0801.0045 [hep-ph]].


\bibitem{Nikolidakis08}
  E.~Nikolidakis, Ph.D. thesis, University of Bern (2008).

\bibitem{Hall:1990ac}
  L.~J.~Hall and L.~Randall,
  ``Weak scale effective supersymmetry,''
  Phys.\ Rev.\ Lett.\  {\bf 65} (1990) 2939.

\bibitem{Ali:1999we}
  A.~Ali and D.~London,
  ``Profiles of the unitarity triangle and CP violating phases in the standard model and supersymmetric theories,''
  Eur.\ Phys.\ J.\ C {\bf 9} (1999) 687
  [hep-ph/9903535].

\bibitem{Isidori:2006qy}
  G.~Isidori, F.~Mescia, P.~Paradisi, C.~Smith and S.~Trine,
  ``Exploring the flavour structure of the MSSM with rare K decays,''
  JHEP {\bf 0608} (2006) 064
  [hep-ph/0604074].


\bibitem{Mercolli:2009ns}
  L.~Mercolli and C.~Smith,
  ``EDM constraints on flavored CP-violating phases,''
  Nucl.\ Phys.\ B {\bf 817} (2009) 1
  [arXiv:0902.1949 [hep-ph]].

\bibitem{Smith:2009hj}
  C.~Smith,
  ``Minimal flavor violation in supersymmetric theories,''
  Acta Phys.\ Polon.\ Supp.\  {\bf 3} (2010) 53
  [arXiv:0909.4444 [hep-ph]].


\bibitem{Paradisi:2008qh}
  P.~Paradisi, M.~Ratz, R.~Schieren and C.~Simonetto,
  ``Running minimal flavor violation,''
  Phys.\ Lett.\ B {\bf 668} (2008) 202
  [arXiv:0805.3989 [hep-ph]].

\bibitem{Colangelo:2008qp}
  G.~Colangelo, E.~Nikolidakis and C.~Smith,
  ``Supersymmetric models with minimal flavour violation and their running,''
  Eur.\ Phys.\ J.\ C {\bf 59} (2009) 75
  [arXiv:0807.0801 [hep-ph]].

\bibitem{oai:arXiv.org:1002.0900}
  G.~Isidori, Y.~Nir and G.~Perez,
  ``Flavor Physics Constraints for Physics Beyond the Standard Model,''  
  Ann.\ Rev.\ Nucl.\ Part.\ Sci.\  {\bf 60} (2010) 355  [arXiv:1002.0900 [hep-ph]].  

\bibitem{Porod:2003um}
  W.~Porod,
  ``SPheno, a program for calculating supersymmetric spectra, SUSY particle decays and SUSY particle production at e+ e- colliders,''
  Comput.\ Phys.\ Commun.\  {\bf 153} (2003) 275
  [hep-ph/0301101].

\bibitem{Porod:2011nf}
  W.~Porod and F.~Staub,
  ``SPheno 3.1: Extensions including flavour, CP-phases and models beyond the MSSM,''
  Comput.\ Phys.\ Commun.\  {\bf 183} (2012) 2458
  [arXiv:1104.1573 [hep-ph]].

\bibitem{SusyFlavor}
  A.~Crivellin, J.~Rosiek, P.~H.~Chankowski, A.~Dedes, S.~Jaeger and P.~Tanedo,
  ``SUSY\_FLAVOR v2: A Computational tool for FCNC and CP-violating processes in the MSSM,''
  Comput.\ Phys.\ Commun.\  {\bf 184} (2013) 1004
  [arXiv:1203.5023 [hep-ph]].

\bibitem{Charles:2004jd}
  J.~Charles {\it et al.}  [CKMfitter Group Collaboration],
  ``CP violation and the CKM matrix: Assessing the impact of the asymmetric $B$ factories,''  
  Eur.\ Phys.\ J.\ C {\bf 41} (2005) 1  [hep-ph/0406184], updated results and plots available at:  http://ckmfitter.in2p3.fr.

\bibitem{Allanach:2001kg}
  B.~C.~Allanach,
  ``SOFTSUSY: a program for calculating supersymmetric spectra,''
  Comput.\ Phys.\ Commun.\  {\bf 143} (2002) 305
  [hep-ph/0104145].

\bibitem{Babu:1987im} 
  K.~S.~Babu,
  ``Renormalization Group Analysis of the {Kobayashi-Maskawa} Matrix,''  
  Z.\ Phys.\ C {\bf 35}, 69 (1987).  

\bibitem{Balzereit:1998id} 
  C.~Balzereit, T. Hansmann, T.~Mannel and B.~Plumper,
  ``The Renormalization group evolution of the CKM matrix,''
  Eur.\ Phys.\ J.\ C {\bf 9}, 197 (1999)  [hep-ph/9810350].  



\end{thebibliography}
\end{document}